\documentclass[12pt]{article}

\usepackage{epsfig}
\usepackage{graphicx}
\usepackage{color}

\newcommand {\hAT} {0.382}

\newcommand {\bi} {\bibitem}
\newcommand {\be} {\begin{equation}}
\newcommand {\beq} {\begin{eqnarray} \nonumber }
\newcommand {\ee} {\end{equation}}

\newcommand{\bq}{\begin{eqnarray}}
\newcommand{\eq}{\end{eqnarray}}

\newcommand{\siml}{\stackrel{<}{\sim}}

\newcommand{\bc}{\begin{center}}
\newcommand{\ec}{\end{center}}

\begin{document}


\title{
Numerical study of the SK Model in magnetic field
}
\author{Alain Billoire and Barbara Coluzzi}

\maketitle

\begin{center}

{\em Service de Physique Th\'eorique} CEA-Saclay \\
Orme des Merisiers  91191 Gif-sur-Yvette France


\end{center}

\vspace{2cm}

\begin{abstract}
\noindent 
We study numerically the Sherrington--Kirkpatrick model as function of
the magnetic field $h$, with fixed temperature $T=0.6 T_c$.  We
investigate the finite size scaling behavior of several quantities,
such as the spin glass susceptibility, looking for numerical evidences
of the transition on the De Almeida Thouless line. We find strong
corrections to scaling which make difficult to locate the transition
point. This shows, in a simple case, the extreme difficulties of spin
glass simulations in non-zero magnetic field.

Next, we study various sum rules (consequences of stochastic stability)
involving overlaps between three and four replicas, which appear to be
numerically well satisfied, and in a non-trivial way.

Finally, we present data on $P(q)$ for a large lattice size ($N=3200$)
at low temperature $T=0.4 T_c$, where, for the first time, the shape
predicted by the RSB solution of the model for non-zero magnetic field is visible.

\end{abstract}

\newpage

\begin{section}{Introduction}
\noindent
The Sherrington--Kirkpatrick (SK) model was introduced some time ago
\cite{ShKi} as a mean field model for spin glasses. Its analytical,
replica symmetry breaking (RSB), solution \cite{Pa} displays in the
glassy phase intriguing features such as an infinite number of pure
states, described by an order parameter which is the non-trivial
probability distribution of the overlap between two of them, $P(q)$.

The applicability of the mean field picture to short range spin
glasses \cite{MaPaRiRuZu} is however still debated, and an alternative
(family of) scenario called the droplet model has been put forward by
several authors.  One may in principle distinguish between these two
theories of finite dimensional spin glasses by looking at the fate of
the glassy phase for non-zero magnetic field.  In the SK model (and
accordingly in the mean field picture) one finds
\cite{MePaVi,BiYo} that a magnetic field (of absolute value) lower than 
the critical De
Almeida Thouless (AT) value $h_{AT}(T)$ \cite{AlTh} 
does not destroy the spin glass ordering, since
the  number of states is still infinite. On the other hand, in the 
droplet picture one has
only two states at $h=0$, related by the global inversion symmetry of
the spins, and  any small magnetic field makes the
system paramagnetic \cite{McMi}-\cite{FiHu}.

It turned out unfortunately \cite{So}-\cite{CrFeJiRuTa} that strong
finite size corrections make difficult to obtain a clear answer from
equilibrium simulations in magnetic field (of the system sizes that
can currently be handled).  Very recently, for example, the authors of a study of the
local excitations of the Edwards--Anderson model (EA)
\cite{LaBoMa} claim that there is no transition whereas results on
the out-of-equilibrium behavior of this model \cite{CrFeJiRuTa} appear
in good agreement with the mean field picture. On the theoretical
side, it has been recently proposed \cite{DeDo} that the transition
below the upper critical dimension $d_u=8$ is governed by a fixed
point different from the AT mean-field fixed point.

This state of affairs motivated us to revisit the case of the SK model
in magnetic field.  There have been indeed few numerical studies of
the SK model in magnetic field at fixed $h$ \cite{CiPaRiRu,PiRi},
\cite{Ri}-\cite{RiSa}, and the behavior of the system as a function of
the magnetic field at fixed $T$ has never been numerically
investigated to our knowledge.

We present results on the spin glass susceptibility and different
dimensionless ratios of $P(q)$ moments, looking for the quantities
that are most appropriate for obtaining numerical evidence of the
transition on the AT line.  We will show in particular that finite
size effects are strongly reduced if one consider the probability
distribution of the absolute value of the overlap, and not the overlap itself.

We moreover consider the overlaps between three and four replicas,
checking the validity of some relations (the so called stochastic
stability sum rules) which are an evident manifestation of the
non-self-averageness of $P(q)$ \cite{YoBrMo,MePaSoToVi} that have been
recently derived under very general properties \cite{Gu2}-\cite{Co}.

Finally we present data for $P(q)$ in magnetic field, where the shape
predicted by the solution of the model, with two peaks separated by a
continuum is visible, for the first time.  All simulations presented
up to now show only a broad  peak around $q_{EA}$. Both large
system sizes (we have $N=3200$), and low temperature (we go down to
$T=0.4$) are needed to see this asymptotic shape.

\end{section}

\begin{section}{Model and Observables}
\noindent
The Sherrington--Kirkpatrick spin glass model with $N$ sites \cite{MePaVi,BiYo} is described 
by the Hamiltonian
\be
{\cal H}_{J}=\sum_{1\le i < j \le N} J_{ij} \sigma_i \sigma_j
-h\sum_{1\le i \le N} \sigma_i,
\ee
here $\sigma_i=\pm 1$ are Ising spins. The sum runs over all spin pairs 
and $J_{ij}$ are quenched identically distributed independent
random variables with mean value $\overline{J_{ij}}=0$ and variance
$1/N$. We take $J_{ij}=\pm N^{-1/2}$.

In order to sample  the probability distribution of the overlap $P(q)$,
one usually considers two independent replicas  $\{ \sigma_i \}$ and 
$\{ \tau_i \}$ evolving contemporaneously and independently:
\begin{eqnarray}
{\cal Q}&=&{1 \over N} \sum_{i=1}^N \sigma_{i} \tau_{i} \\
P(q)&\equiv&\overline{P_J(q)}\equiv
\overline{ \langle \delta(q-{\cal Q})\rangle},
\end{eqnarray}
where the thermal average $\langle \cdot \rangle$ corresponds to the
time average during the simulation, whereas $\overline{(\cdot )}$
stands for the average over the $J_{ij}$ realizations.  This is the
order parameter of the model, which in the thermodynamic limit
behaves like
\begin{equation}
P(q)=\left \{
\begin{array}{lcl}
\delta(q-q_{EA}) & \hspace{.3cm} & |h|>h_{AT}(T) \\
x_m \delta(q-q_m) +\tilde{P}(q)+ x_M \delta(q-q_{EA}) 
& \hspace{.3cm} & 0<|h|<h_{AT}(T) \\
\frac{1}{2} \left [\tilde{P}(q)+\tilde{P}(-q) \right ] +
{x_M \over 2} \left [ \delta (q-q_{EA}) + \delta (q+q_{EA}) \right ]
& \hspace{.3cm}& h=0, T<T_c
\end{array}
\right .
\label{pq}
\end{equation}
where $h_{AT}(T)$ is the critical value of the magnetic field on the
AT line, with $h_{AT}(T) \sim (4/3)^{1/2}(T_c-T)^{3/2}$ for $T
\rightarrow T_c^-$ ($T_c=1$ in this model) \cite{AlTh}.  For $T
\rightarrow T_c^-$ one finds that $x_m \propto q_m \propto h^{2/3}$, and
$(q_{EA}-q_m)
\propto (x_M-x_m) \propto (h_{AT}(T) -h)$.
Note that at $h=0$ the function $P(q)$ is symmetric, reflecting the
global flip $\{ \sigma_i \} \rightarrow \{- \sigma_i \}$
symmetry of the system, and the $\delta$-function at $q_m$
disappears. As we are going to discuss in detail, this implies a
singular behavior for different quantities in the $h \rightarrow 0$
limit, which is among the main sources of difficulties in finding
evidence for the phase transition in magnetic field.

In order to locate numerically phase transitions, in our case the AT
line, it is common practice to consider dimensionless ratios of
moments of $P(q)$, which are expected to have a fixed point (with
respect to $N$) at the critical point, like the (connected) Binder parameter
\cite{Bi,BhYo}
\be
B(h,T)= {1 \over 2} \left ( 3 - {\overline{\langle (q-\overline{\langle q \rangle})^4 
\rangle} \over
\overline{\langle (q-\overline{\langle q \rangle})^2 \rangle}^2} \right ),
\label{binder}
\ee
and the skewness
\be
S(h,T)= {\overline{\langle (q-\overline{\langle q \rangle})^3 \rangle} \over
\overline{\langle (q-\overline{\langle q \rangle})^2 \rangle}^{3/2}}.
\label{skewness}
\ee
Here $B(h,T)$ is defined in such a way that it is zero for a Gaussian
distribution and one for a two equal weights $\delta$-function
distribution, whereas the skewness is a measure of $P(q)$ asymmetry,
which is non-zero in our case of non-zero magnetic field.

Further evidence for the presence of a phase transition should 
come from the behavior of the spin glass susceptibility
\be
\chi_{SG}(h,N)\propto N 
\left ( \overline{ \langle q^2 \rangle} - 
\overline{\langle q \rangle}^2 \right ),
\ee
which is expected to diverge, 
in the thermodynamic limit, when entering in the spin glass phase, like
\be
\chi_{SG}(h,\infty) \propto \frac{1}{h-h_{AT}} \hspace{.3in} h \rightarrow h^+_{AT},
\hspace{.3in} T \ {\rm fixed}.
\ee
According to usual finite size scaling arguments, the finite size behavior is
(for $h$ near to $h_{AT}$): 
\be
\chi_{SG}(h,N)= N^{1/3} \tilde{\chi}_{SG} \left [ N^{1/3} (h-h_{AT}) \right ],
\hspace{.3in} T \ {\rm fixed}.
\ee
The finite size behavior of the spin glass susceptibility in the SK
model on the AT line was numerically studied in \cite{CiPaRiRu}, and
the scaling $\approx N^{1/3}$ was checked.

The non-self-averageness of $P(q)$ is among the many fascinating
features of the SK model.  For instance, considering four
replicas of the system, one finds \cite{MePaSoToVi} that
$\overline{P_J(q_{12})P_J(q_{34})}\neq P(q_{12})P(q_{34})$, whereas
the following relations hold:
\begin{eqnarray}
P(q_{12},q_{34}) & \equiv & \overline{P_J(q_{12})P_J(q_{34})} =
{2\over 3} P(q_{12})P(q_{34})+{1\over 3}P(q_{12}) \delta(q_{12}-q_{34}) \label{sumrule1} \\
P(q_{12},q_{13}) & \equiv &  \overline{P_J(q_{12})P_J(q_{13})} =
{1\over 2} P(q_{12})P(q_{13})+{1\over 2}P(q_{12}) \delta(q_{12}-q_{13}).
\label{sumrule2}
\end{eqnarray}
These kind of relations, which are non-trivially verified if the replica
symmetry is  broken ($P(q)\neq\delta(q-q_{EA})$), were recently derived under 
very general conditions \cite{Gu2}-\cite{Co}, such as stochastic stability 
\cite{Pa2}. Infinitely many  sum rules  follow. We consider in particular the relations:
\begin{eqnarray}
R^a_{1234}(h,T)&=&\overline{\langle q_{12}q_{34} \rangle}-{2\over
3}\overline{\langle q \rangle}^2- {1\over 3} \overline{\langle q^2
\rangle}=0 \label{Ra1234}
\\ 
R^b_{1234}(h,T)&=&\frac{\overline{\langle q_{12}q_{34}\rangle}-
\overline{\langle q \rangle}^2}{\overline{\langle q^2 \rangle}-\overline{\langle q \rangle}^2}=
\frac{{\chi_{SG}^{1234}(h,T)}}{\chi_{SG}(h,T)}={1\over 3}\\
R^a_{1213}(h,T)&=&\overline{\langle q_{12}q_{13} \rangle}-{1\over 2}\overline{\langle q \rangle}^2-
{1\over 2}\overline{\langle q^2 \rangle}=0\\
R^b_{1213}(h,T)&=&\frac {\overline{\langle q_{12}q_{13} \rangle}-\overline{\langle q \rangle}^2 }
{\overline{\langle q^2 \rangle}-\overline{\langle q \rangle}^2}=\frac{{\chi_{SG}^{1213}(h,T)}}
{\chi_{SG}(h,T)}=
{1\over2}\\
R^2_{1234}(h,T)&=&\overline{\langle q^2_{12}q^2_{34} \rangle}-{2\over 3}\overline{\langle q^2 \rangle}^2-
{1\over 3}\overline{\langle q^4 \rangle}=0\\
R^2_{1213}(h,T)&=&\overline{\langle q^2_{12}q^2_{13} \rangle}-{1\over 2}\overline{\langle q^2 \rangle}^2-
{1\over 2}\overline{\langle q^4 \rangle}=0
\label{sumrules}
\end{eqnarray}
Relations $R_{1234}^{a,b}(h,T)$ and $R_{1213}^{a,b}(h,T)$, which have
to our knowledge never been previously investigated numerically, are
expected to be verified only at non-zero magnetic field, since these
relations are not invariant under a global flip, and an infinitesimal
magnetic field was implicit in the derivation of equations
(\ref{sumrule1}-\ref{sumrule2}). On the other hand, relations
$R^2_{1234}(h,T)$ and $R^2_{1213}(h,T)$ are valid for zero magnetic
field also, and were already studied (for $3d$ and $4d$ Ising spin
glasses at zero magnetic field \cite{MaPaRiRuZu,MaPaRuRi,MaZu}).

We measured also the following ratios of moments which are non-zero
when the system is non-self-averaging, and have been introduced  for
locating the transition point
\cite{MaNaZuPaPiRi,RiSa}: 
\begin{eqnarray}
G(h,T)&=&
\frac{\overline{\langle q^2 \rangle^2}-\overline{\langle q^2\rangle}^2}
{\overline{\langle q^4 \rangle}-\overline{\langle q^2\rangle}^2} \\
G_c(h,T)&=&
\frac{\overline{\langle (q-\langle q \rangle)^2 \rangle^2}-
\overline{\langle (q -\langle q \rangle)^2\rangle}^2}
{\overline{\langle (q-\langle q \rangle)^4 \rangle}-
\overline{\langle (q-\langle q \rangle)^2 \rangle}^2} \\
A(h,T)&=&
\frac{\overline{\langle q^2 \rangle^2}-
\overline{\langle q^2\rangle}^2}
{\overline{\langle q^2 \rangle}^2} \\
A_c(h,T)&=&
\frac{\overline{\langle (q-\langle q \rangle)^2 \rangle^2}-
\overline{\langle (q -\langle q \rangle)^2\rangle}^2}
{\overline{\langle (q-\langle q \rangle)^2 \rangle}^2} 
\label{Ac}
\end{eqnarray} 

In the infinite volume limit, $G(h,T)$ is expected to take the
constant value $1/3$ in the glassy phase, because of relation
(\ref{sumrule1}), whereas $G_c(h,T)$, $A(h,T)$ and $A_c(h,T)$ are
non-trivial function of $h$ and $T$, that are zero in the whole
paramagnetic phase.

These parameters should be particularly useful when the Binder
parameter behaves non monotonically, taking both positive and negative
values, as it is found to happen when there is no time-reversal
symmetry in the Hamiltonian \cite{CiPaRiRu,PiRi1,MaNaZu,PaPiRi,HuKa}
(like in our case of non-zero magnetic field) and in systems where the
mean field solution is one step replica symmetry breaking
\cite{CaCoPa}. Their behavior has been extensively studied and they
have been  applied to a number of models \cite{MaNaZuPaPiRi,RiSa,
PaPiRi,HuKa,DrBoMo,BaCrFeMaPeRuTaTeUlUn}.

The study of $A$ and $A_c$ allows to check if the numerator in $G$ and
$G_c$ is really non zero or if it is approaching zero for increasing
volumes together with (or more slowly than) the denominator
\cite{MaNaZuPaPiRi,RiSa}. This is obviously not expected to happen in
the glassy phase of the SK model. As a matter of fact, it was shown
\cite{RiSa} that $G$ and $G_c$ should take the universal values 1/3
and 13/31 respectively at zero temperature for any finite volume Ising
system under quite general hypothesis, i.e. even if the order
parameter is self-averaging.

The connected parameter $G_c$ should be the most effective quantity to look 
at for locating the transition point, since it seems to be the one less 
affected by finite size corrections to scaling \cite{RiSa}. 

Relations $R^b_{1234}$, $R^b_{1213}$ should behave as $G$ and are in
principle good candidates for obtaining evidence of the
transition. Nevertheless we will show that their finite size behavior,
as well as the one of $G$ itself, is definitely different from that of
$G_c$ (at least for the considered $N$ values) and that they do not
help to locate the transition point.

The main source of finite size effects is the global reversal of all
spins.  It does not occur in the thermodynamic limit, but when $h$ is
exactly zero.  It  does occur however  in a finite volume, and as a
consequence $P(q)$ develops a tail in the $q<0$ region.  This tail
is significant \cite{PiRi,BiCo} even for a size as large as
$N=1024$ and a magnetic field value $h=0.3$ (at temperature $T=0.6$).
This was observed also in finite dimensional systems
\cite{CaPaPaSo,CiPaRiRu,PiRi,MaNaZu} and it is expected to strongly affect
the scaling of different quantities. In order to reduce
its importance, one may use \cite{BaCiPaRiPeRu,CiPaRiRu}
the ``absolute'' spin glass susceptibility defined as 

\be
\chi_{SG}^{abs}(h,N)=N \left ( \overline{\langle q^2 \rangle} -
\overline{\langle | q | \rangle}^2 \right ).
\ee

More generally, one can define ``absolute'' variants of all quantities
defined in Equations (\ref{binder}-\ref{skewness}),(\ref{Ra1234}-\ref{Ac}). 
In the following we will
systematically study the differences between usual quantities and
``absolute'' ones (that will be labeled by the superscript $abs$),
trying to clarify which are the most appropriate to look at in order
to get evidence for the transition.
\end{section}

\begin{section}{Simulations}
We fixed the temperature at $T=0.6$, where the AT line corresponds to
the critical value $h_{AT}(T=0.6)\simeq 0.382$ \cite{CrRi}.  We use
the magnetic Parallel Tempering algorithm (h-PT), described in details
in our previous paper \cite{BiCo}. We consider $n=49$ replicas, each
one at a different magnetic field $h$ from a set of $n$ different
values both within and without the AT lines, from $h_{min}=-0.6$ to
$h_{max}=0.6$ at equally spaced intervals of $\delta
h=0.025$. Exchange of $h$ values between nearest neighbor replicas are
allowed with the usual Monte Carlo acceptance probability. Moreover
the sign of all spins of a replica at $h=0$ can be reversed with
probability 1/2. This makes easier the passage from negative to
positive $h$ values and vice-versa.

This h-PT algorithm is ideally suited to obtain the behavior of quantities
as function of the magnetic field at fixed temperature. However it was
found \cite{BiCo} that its efficiency rapidly decreases when
simulating large systems, most probably because of chaos with
magnetic field. At variance with the case of temperature chaos, the
effect of chaos with magnetic field becomes evident already on a size of
order $N=1024$ and $\delta h \approx 0.15$. This means that the phase
spaces explored by the system at equilibrium at $h$ and $h+\delta h $
become quite different when $\delta h \approx 0.15$.  Therefore
$N=1024$ is the largest size we could thermalize with this method.

We perform 50000+50000, 100000+100000 and 300000+300000 h-PT
sweeps for $N=64$, $256$ and $1024$ respectively, the first half of
each run being discarded from the statistics.  Thermalization was
checked by comparing the data obtained in the second part with the
ones of the second quarter.  We simulated four sets of replicas
evolving contemporaneously and independently (i.e. $49 \times 4=196$
replicas). Data are averaged over $256$ disorder configurations for
each system size. Statistical errors are evaluated from (disorder)
sample-to-sample fluctuations by using the Jack-knife method.

Looking for evidences of the shape of $P(q)$ with $h\neq0$
predicted by the solution of the model (Equations (\ref{pq})),
we performed a large scale  simulation for a system of $N=3200$
spins, with the usual PT in temperature algorithm, taking $n=38$ equally 
spaced temperature values between $T_{min}=0.4$ and $T_{max}=1.325$, 
at magnetic field $h=0.3$. Results were averaged over
128 different disorder realization and we performed $400000+400000$ PT steps
for each sample, checking thermalization by comparing the $P(q)$
obtained in the second half and in the second quarter of the run.
\end{section}

\begin{section}{Results and Discussion}

\begin{subsection}{Energy, magnetization and magnetic susceptibility}
\noindent
We plot in [Fig. 1]  the energy density and  the 
magnetization as a function of $h$ for the different sizes considered. 
In the same figure, we also present
data on   the mean overlap 
$\overline{\langle q \rangle}$ and on the mean absolute value of the 
overlap $\overline{\langle | q | \rangle}$. It is evident from these data
that the two quantities definitely differ for $h$  as large as 
$h \simeq 0.4$ (i.e. larger than $h_{AT}$) for $N=64$ and up to $h \simeq 0.2$
for $N=1024$.

\begin{figure}[htbp]
\begin{center}
\leavevmode
\epsfig{figure=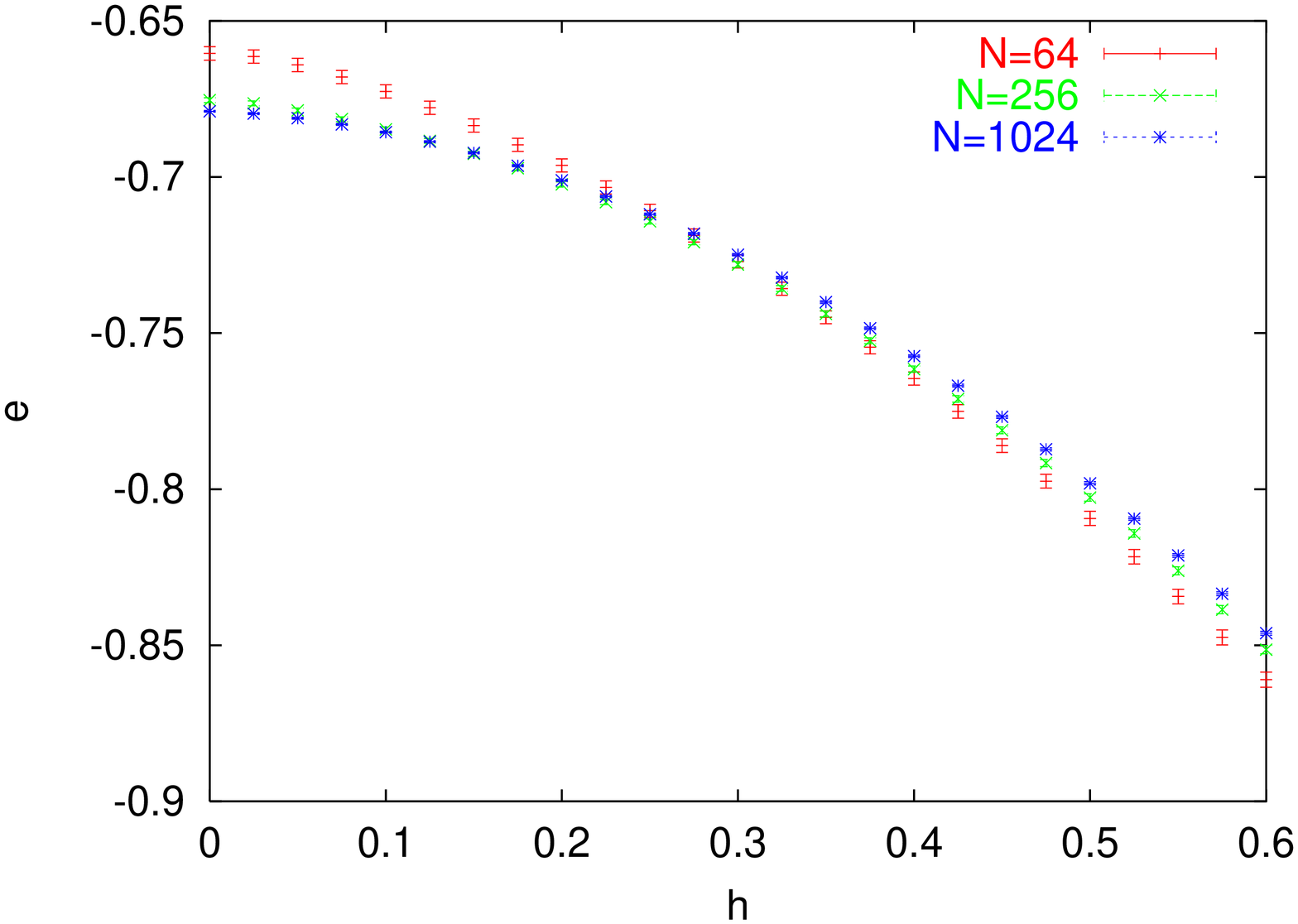,angle=270,width=8cm}
\epsfig{figure=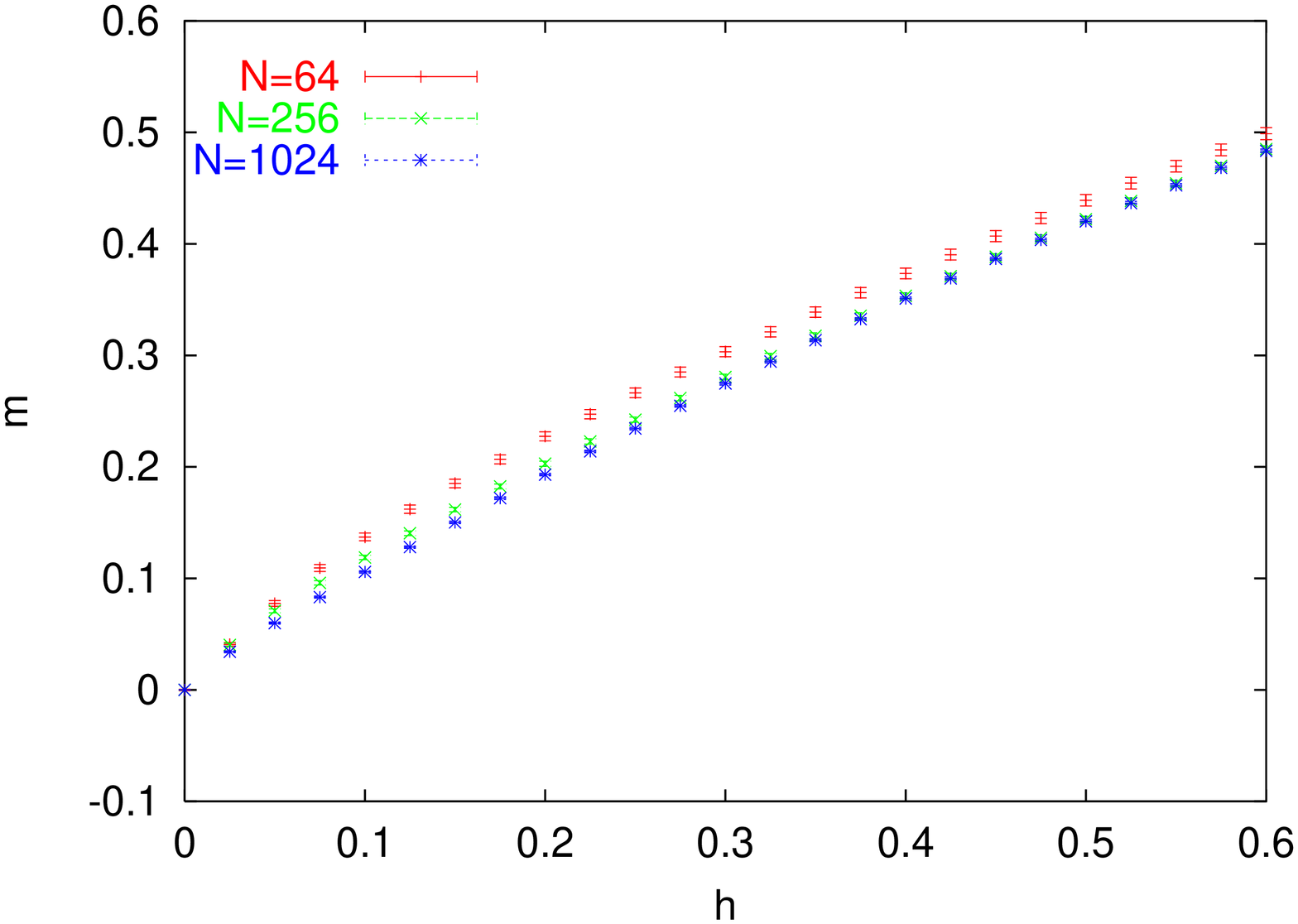,angle=270,width=8cm}
\epsfig{figure=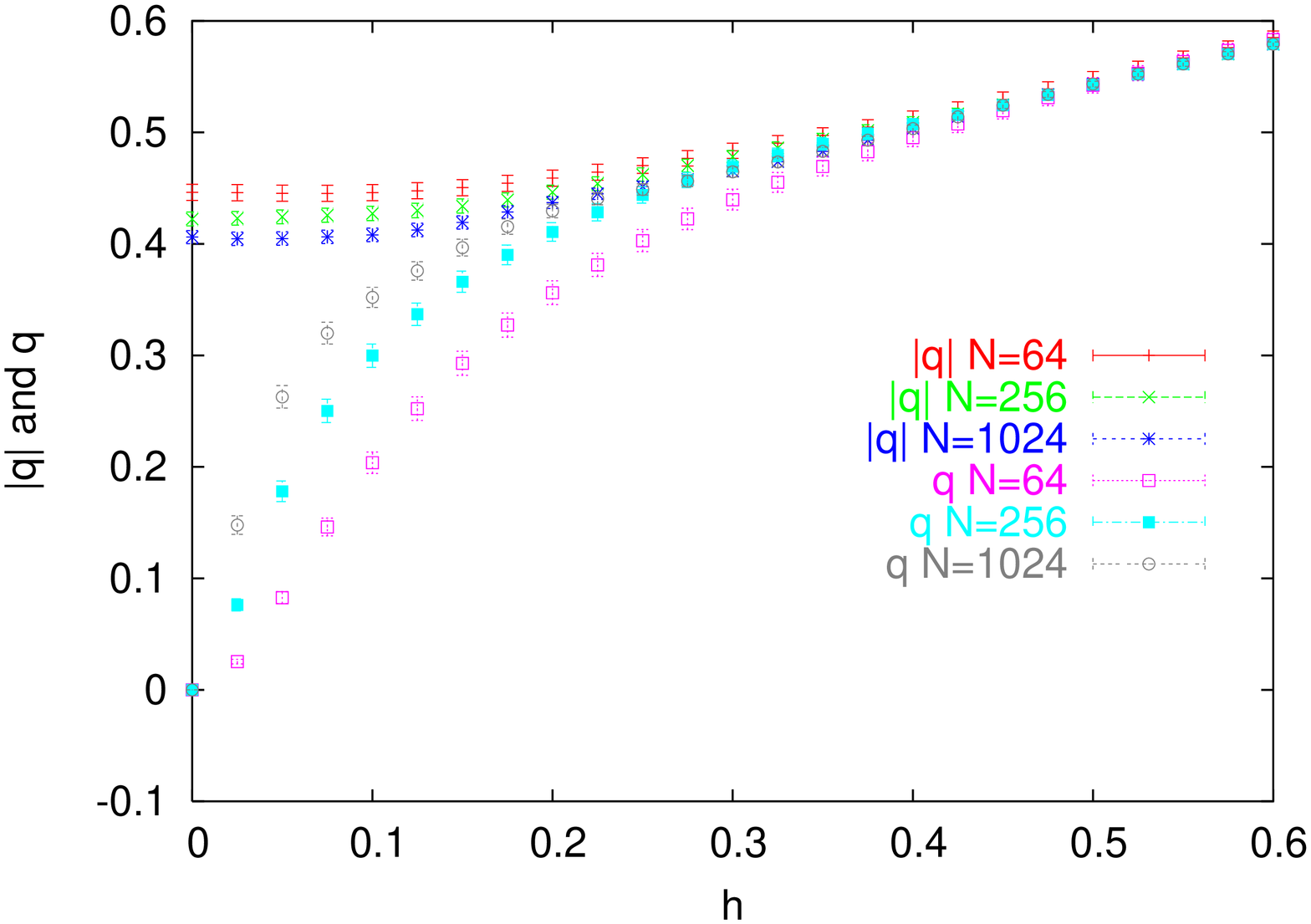,angle=270,width=12cm}
\caption{The energy density (top left), the magnetization density 
(top right) and the mean value of the overlap : $\overline{\langle q
\rangle}$ and $\overline{\langle |q| \rangle}$ (bottom) at $T=0.6$, 
as a function of the magnetic field, for the different system sizes.}
\end{center}
\end{figure}
\begin{figure}[htbp]
\begin{center}
\leavevmode
\epsfig{figure=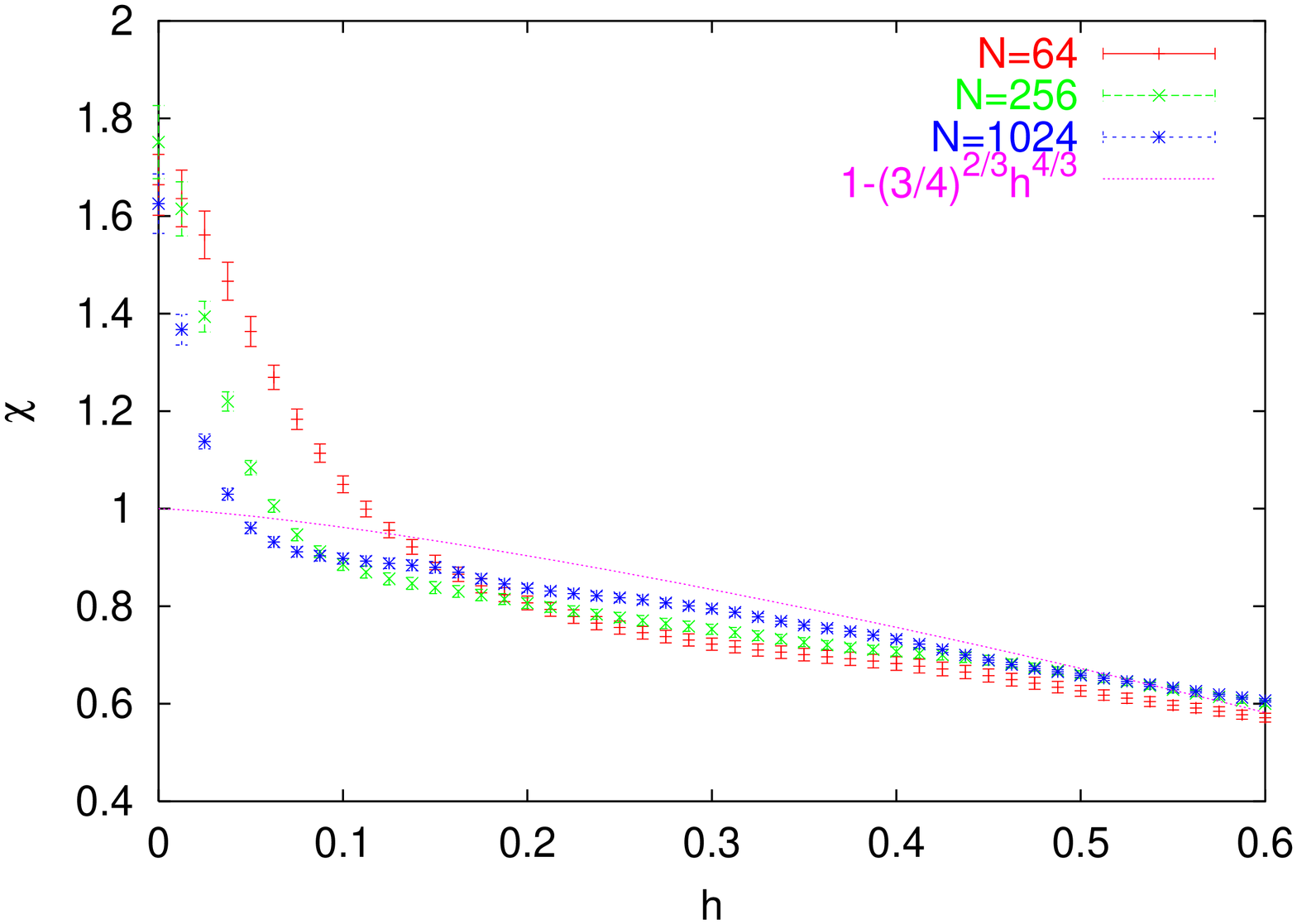,angle=270,width=8cm}
\epsfig{figure=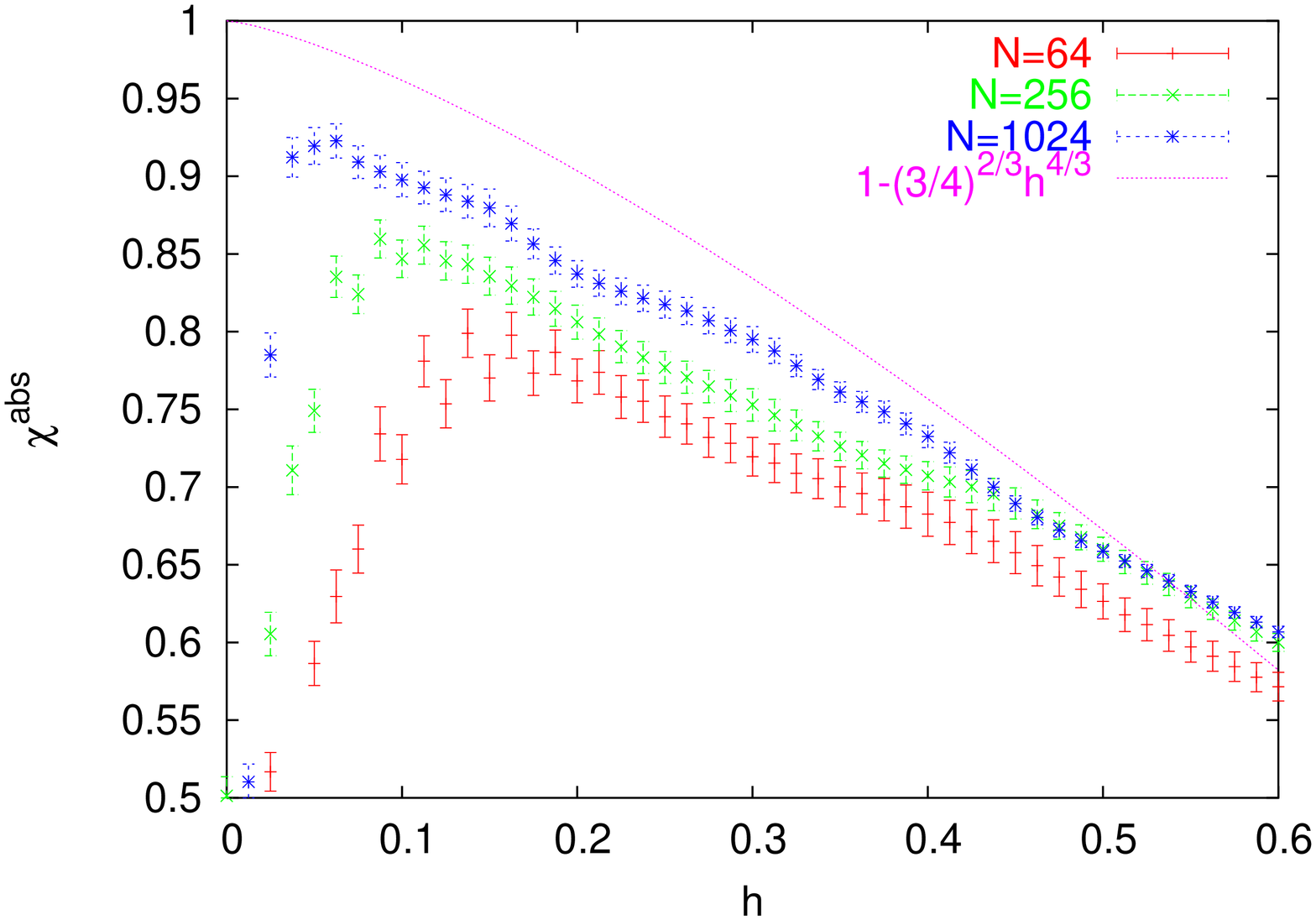,angle=270,width=8cm}
\caption{On the left:  usual magnetic susceptibility at $T=0.6$, 
as a function of the magnetic field, for the different values of the system 
size, compared with the predicted infinite volume analytic behavior 
(see text). On the right:  the magnetic susceptibility from the
absolute value of the magnetization.
In both plots the two estimates of the susceptibility (\ref{chi1}) and (\ref{chi2}) are  plotted.}
\end{center}
\end{figure}

In  [Fig. 2] we plot the magnetic susceptibility computed as
\be 
\chi(h,N)= {\partial \overline{\langle m \rangle} \over \partial h},
\label{chi1}
\ee
and as
\be
\chi(h,N)=N \beta( \overline{\langle m^2 \rangle}- \overline{\langle m \rangle^2}).
\label{chi2}
\ee
Excellent agreement\footnote{Although formula \ref{chi1} and
\ref{chi2} are identical mathematically, their Monte Carlo estimates
can disagree, if the sampling is bad, or if $\delta h$ (we use finite
difference to estimate
\ref{chi1}) is too large.} is observed between the two
estimates, showing that good sampling has been achieved. We also plot
the analytical estimate \cite{Pa3} (obtained in the so-called Parisi  approximation 
for the Landau potential, and
for $T \rightarrow T_c^-$):
\be
\chi(h,\infty)=1-(3/4)^{2/3}h^{4/3},
\label{chi_ana}
\ee 
which is in quite good agreement with the data for $h$ not too
small\footnote{A different formula appears in \cite{PaTou} and e.g. \cite{CrRiTe}, with the 
order $h^{4/3}$ term multiplied by $7/3$.}. In the
thermodynamic limit, $\chi=\beta(1-\overline{\langle q
\rangle})=\beta/\beta=1$ in the whole spin glass phase for
$h\rightarrow 0$. On a finite  system however one must take into
account the symmetry with respect to the flip of all the spins and
$\chi(h=0,\infty)=\beta(1-\overline{\langle q
\rangle})=\beta\simeq 1.666$. Our data for $\chi$ show clearly the crossover between
these $h=0$ and $h\neq 0$ regimes. Figure 2 shows that finite size
effects are positive for low $h$, and negative for larger $h$. This
change of sign must happen since the susceptibility has sizable finite
size corrections whereas the magnetization is (trivially) size
independent for $h=0$, and very weakly size dependent for large $h$.

In order to better clarify the situation, we consider also the
``absolute'' magnetic susceptibility $\chi^{abs}$ (see the same figure
on the right), obtained from the probability distribution of the
absolute value of the magnetization.  Note that for the ``absolute''
susceptibility the estimates of equations (\ref{chi1}) and
(\ref{chi2}) do not coincide, but when finite size effects are
negligible. \end{subsection}

\begin{subsection}{On the critical behavior at the AT line}
\noindent
Let us now look at the dimensionless ratios of $P(q)$ moments which
should intersect at $h_{AT}(T)$ giving evidences for the phase
transition (Here $h_{AT}\simeq\hAT$). We present in [Fig. 3] the behavior
of the Binder parameter and of the skewness. The original $B$ and $S$
of Equation (\ref{binder}) and (\ref{skewness}) on the left of the
figure, the ``absolute'' variants on the right.  We see that in all
cases it would be a hard task to get a clear unambiguous determination
of the critical point from the data. At variance with the behavior of
the $h=0$ Binder parameter which is always positive and increases
continuously for decreasing $T$'s, all four quantities in [Fig 3]
(the absolute value does not help)
display a non-monotonic behavior, taking negative values on a large
part of the interval.  For $B$ and $S$, the intersection between
$N=256$ and $N=1024$ happens at a definitely too small $h
\simeq 0.2$, showing the presence of strong finite size scaling
corrections (In all cases the value $N=64$ turns out to be too small
to give interesting results).  The effect of the absolute value is
strong, particularly for $N=64$ and $256$.  It goes in the right
direction since comparing data for $N=64$, $256$ and $1024$, we now
find a monotonous crossing point behavior which approach $h_{AT}$ from
above, with the intersection between $N=256$ and $1024$ data happening
(within the error) at the correct value $h_{AT}$, the agreement being
best in the case of the skewness.

\begin{figure}[htbp]
\begin{center}
\leavevmode
\epsfig{figure=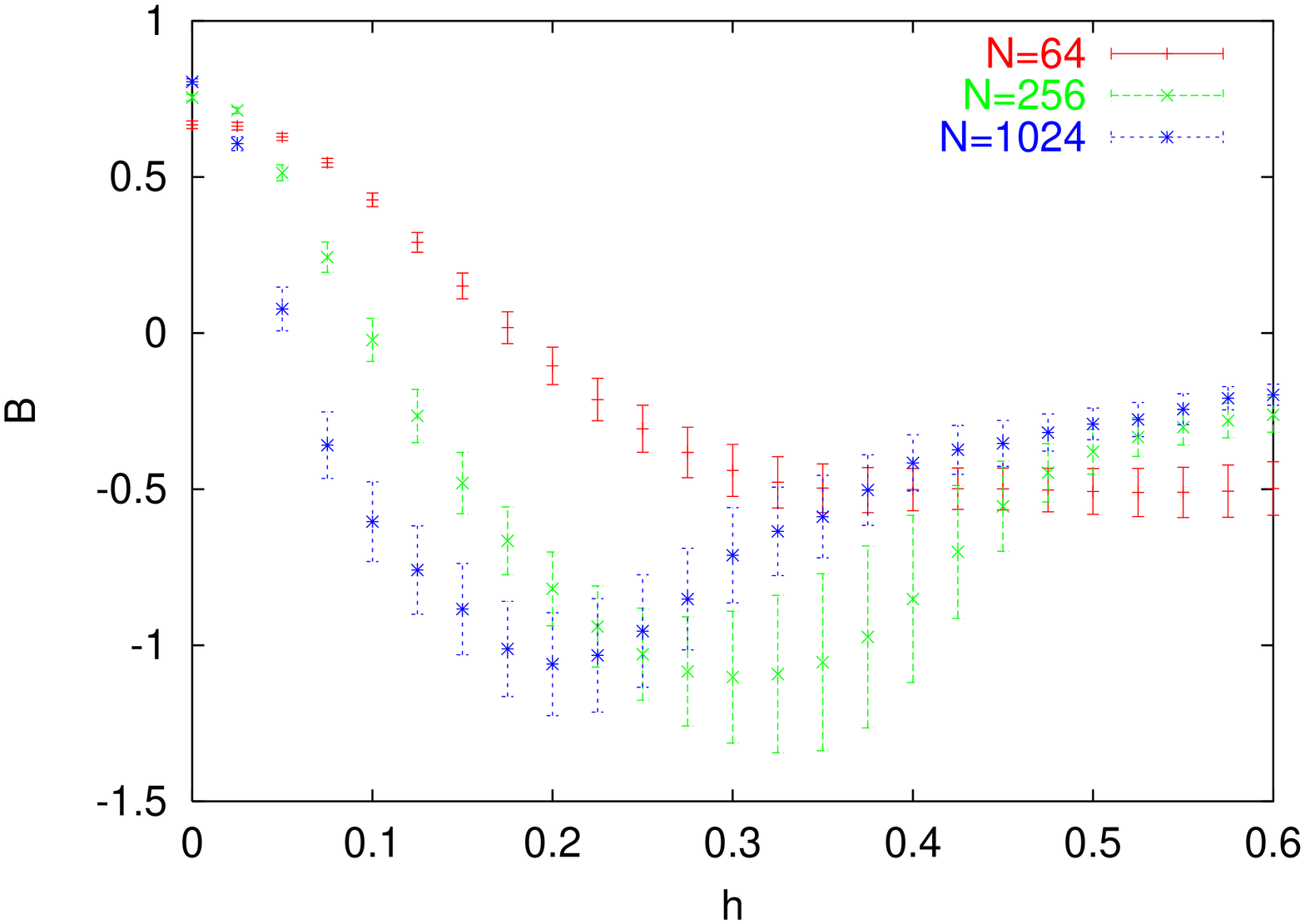,angle=270,width=8cm}
\epsfig{figure=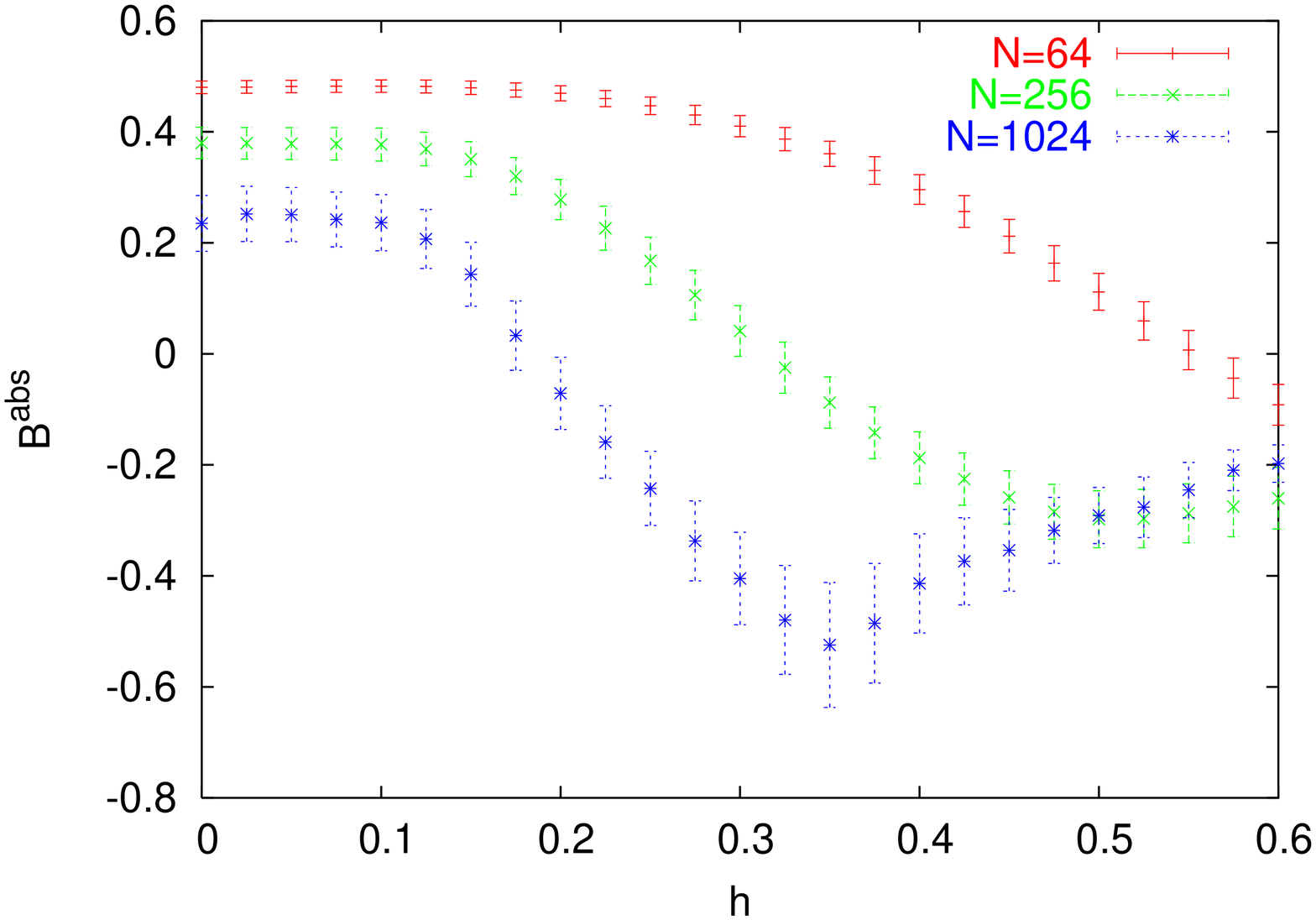,angle=270,width=8cm}
\epsfig{figure=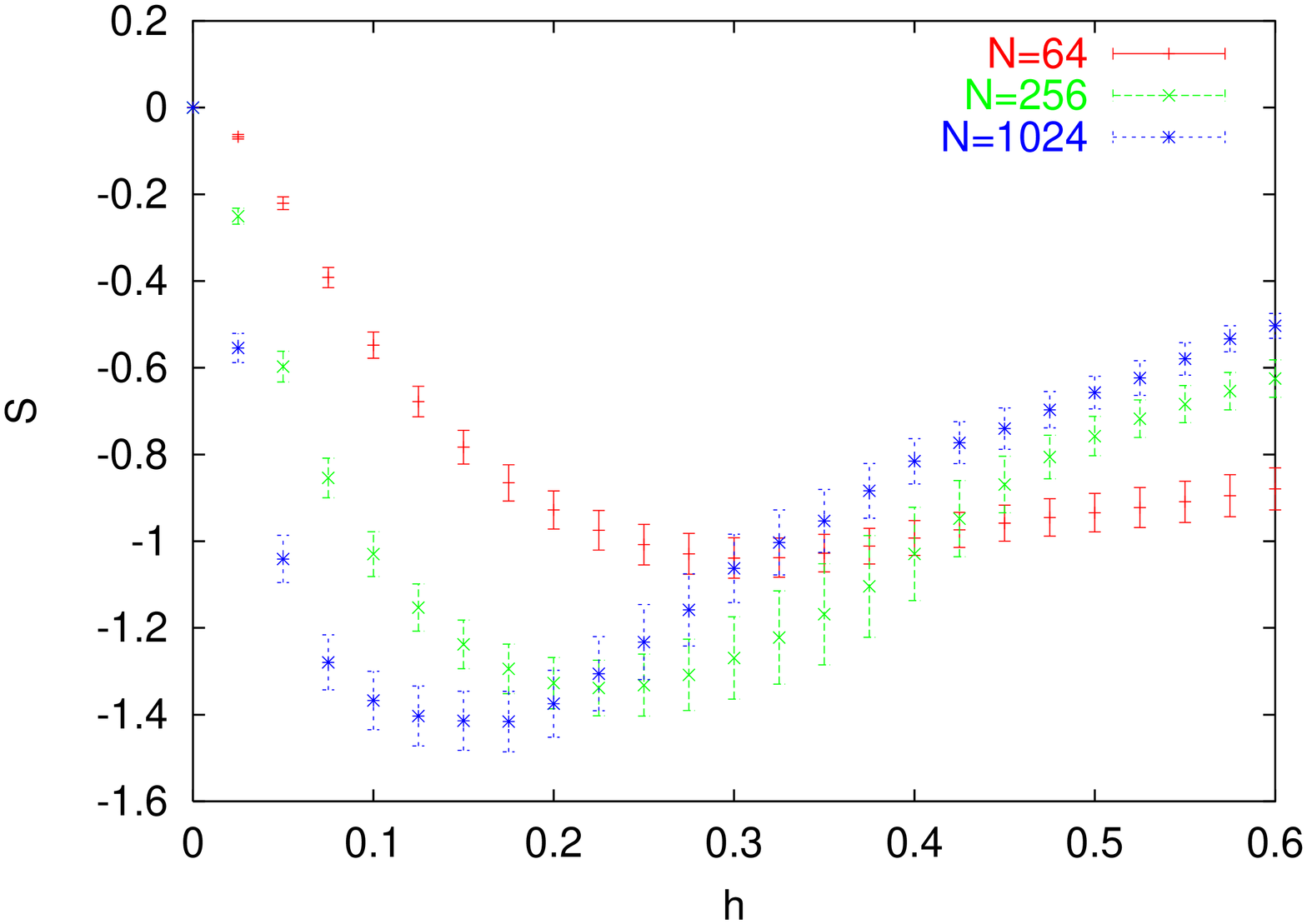,angle=270,width=8cm}
\epsfig{figure=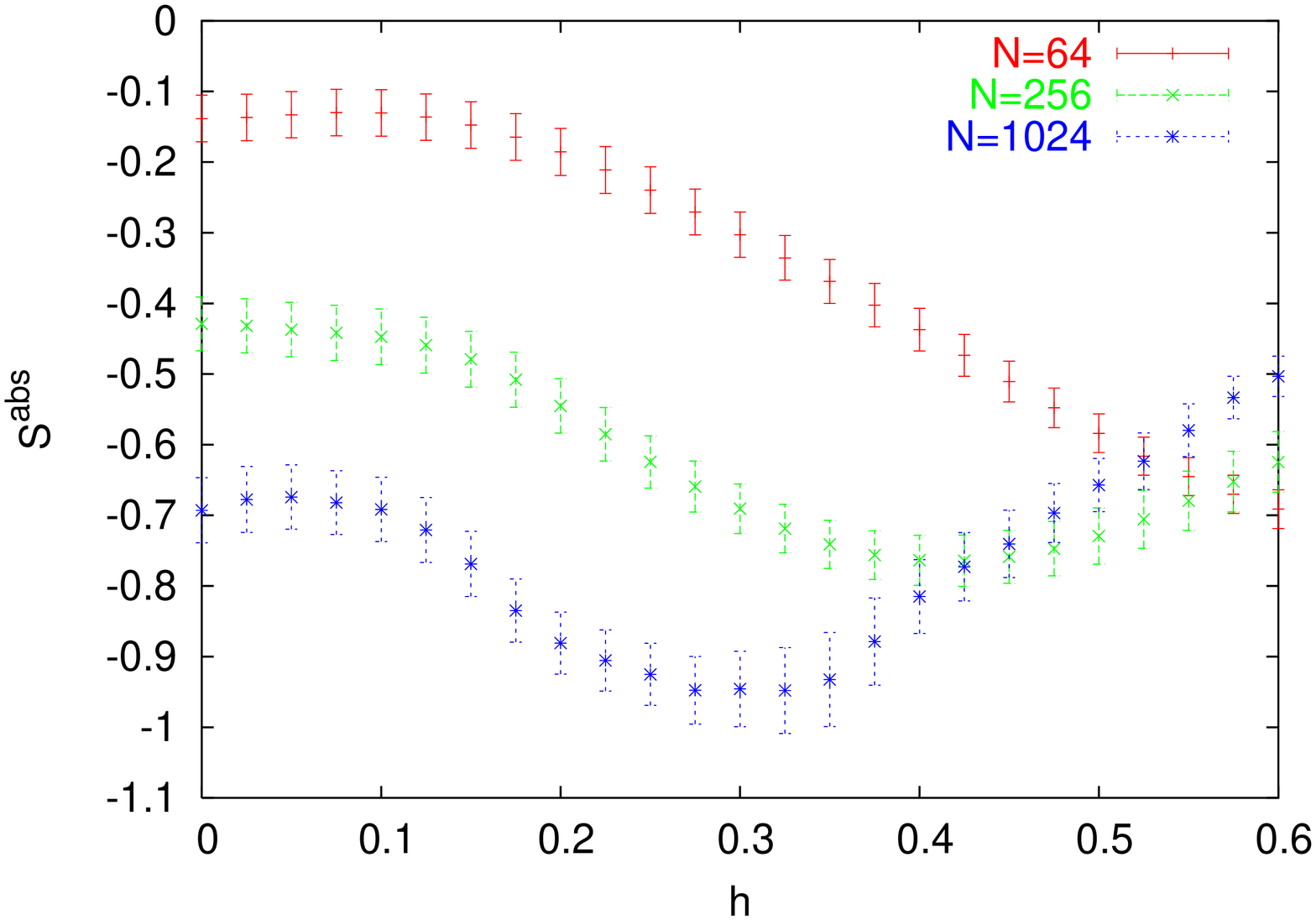,angle=270,width=8cm}
\caption{The behavior of the Binder parameter $B(h,T)$ (up, left) and of the 
skewness $S(h,T)$ (bottom, left) at $T=0.6$, as a function of the
magnetic field for the different system sizes. Here $h_{AT}\simeq\hAT$. On
the right are plotted the corresponding quantities $B^{abs}(h,T)$ (up,
right) and $S^{abs}(h,T)$ (bottom, right) obtained from the
distribution of the absolute values of the overlap.}
\end{center}
\end{figure}

Next we consider in [Fig. 4] the parameters based on $P(q)$ glassy
phase non-self-averageness. The first observation is that in order to
obtain some information one has to look at connected quantities, since
$G$ and $A$ have a definitely different behavior from the others and
do not cross as one could expect.  More in detail, curves for $G$
corresponding to $N=64$ and $256$ are nearly constant (within the
errors) in the whole considered $h$-range. For $N=1024$, $G$ is in
agreement within errors with the thermodynamic limit value $1/3$ in
the glassy phase, and is smaller in the paramagnetic phase.  For $h$
as large as $0.6$ one still finds $G\simeq 0.28$.
In the case
of the EA model at fixed magnetic field the parameter $G$ was found
\cite{RiSa} less appropriate than $G_c$. It decreases more slowly when
entering in the paramagnetic region and curves for different sizes
cross at definitely too large temperatures (here quite small sizes
with $N$ between 5 and 64 were considered).

Curves for $A$ are monotonic and
approach zero (though with strong finite size corrections) at large
$h$, making evident that  $P(q)$ is self-averaging in the
paramagnetic phase, but they do not cross correctly and in particular
at $h=0$ one finds $A(N=1024)\simeq A(N=256)<A(N=64)$.

The connected parameters display a more interesting behavior and they do
indeed cross. We  find a qualitative similar behavior near the transition
point for $G_c$ and $A_c$. Our statistical errors are quite 
large and do not allow a precise determination of the crossing points, 
nevertheless curves for the two smaller sizes seem to roughly intersect at 
the correct value $h \simeq 0.4 \simeq h_{AT}$, whereas data for $N=256$ and 
$1024$ appear to cross at a lower $h$-value, this being more evident in the 
case of $A_c$.
The behavior seems still better when looking at the corresponding
``absolute'' parameters. In particular, there is no irregular behavior
for $h \rightarrow 0$ and the statistical errors are definitely smaller,
allowing a more precise evaluation of the crossing points. Curves
corresponding to $N=64$ and $256$ intersect at $h \simeq 0.6$ whereas
$N=256$ and $1024$ are clearly crossing at the right value, $h \simeq
0.4 \simeq h_{AT}$.

To conclude the discussion on these parameters, we note that although
$G_c$ and $A_c$ are certainly interesting to look at for getting
evidences of the transition (and of non-self-averaging), one should
not overlook their large statistical fluctuations, much larger than the fluctuations
of the usual Binder parameter or the skewness, therefore a larger number of
samples would be required in order to obtain precise measures.

\begin{figure}[htbp]
\begin{center}
\leavevmode
\epsfig{figure=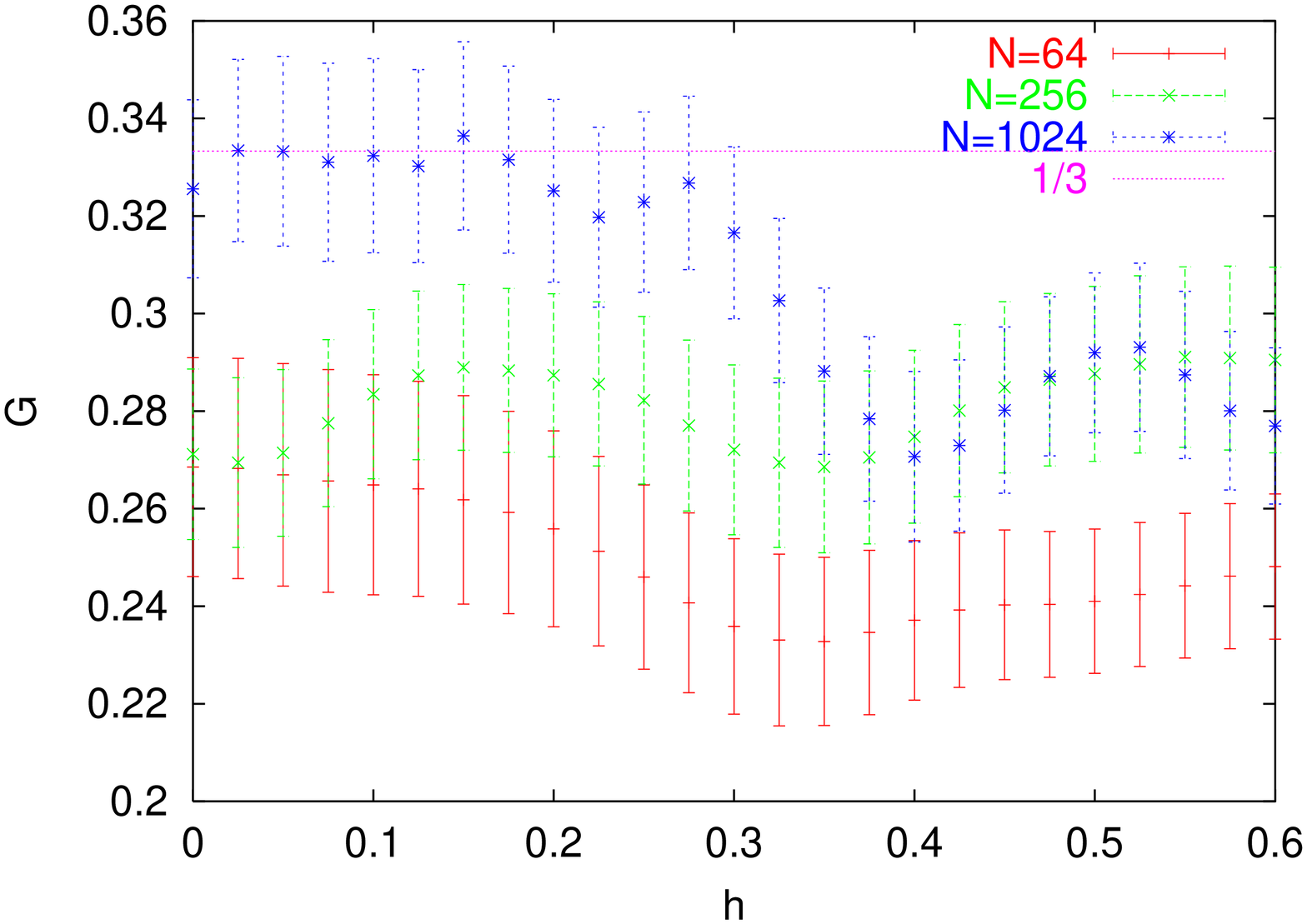,angle=270,width=8cm}
\epsfig{figure=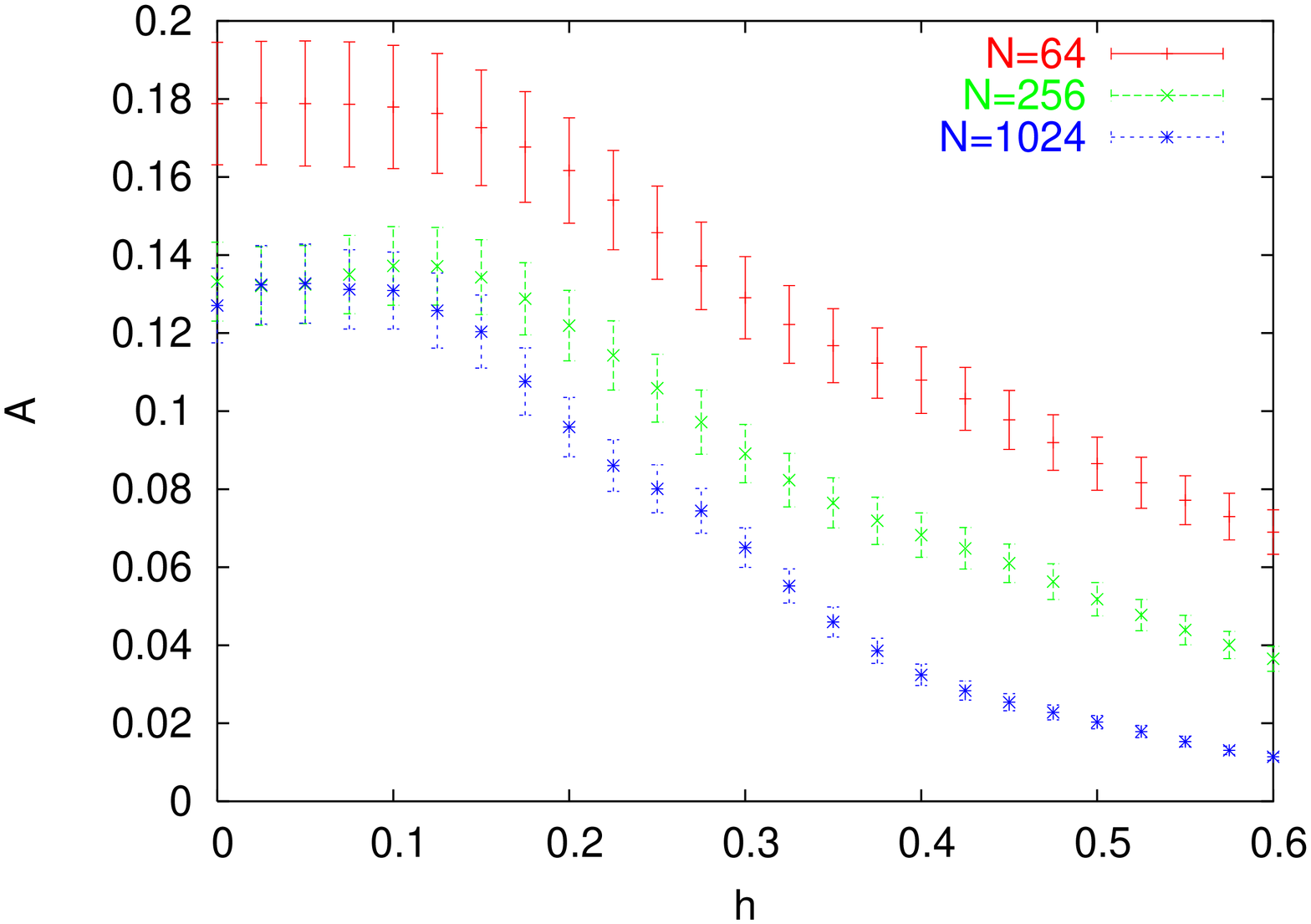,angle=270,width=8cm}
\epsfig{figure=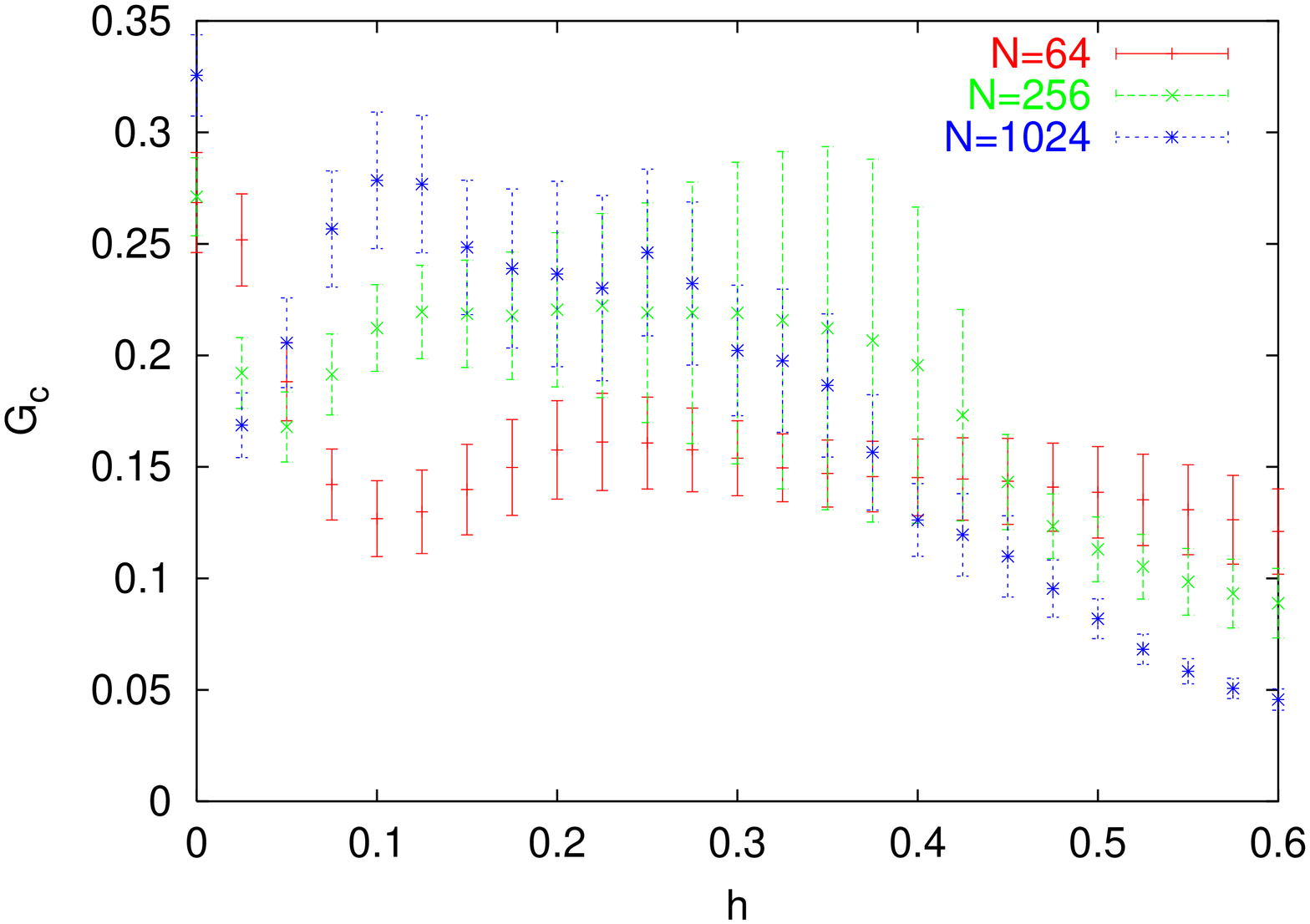,angle=270,width=8cm}
\epsfig{figure=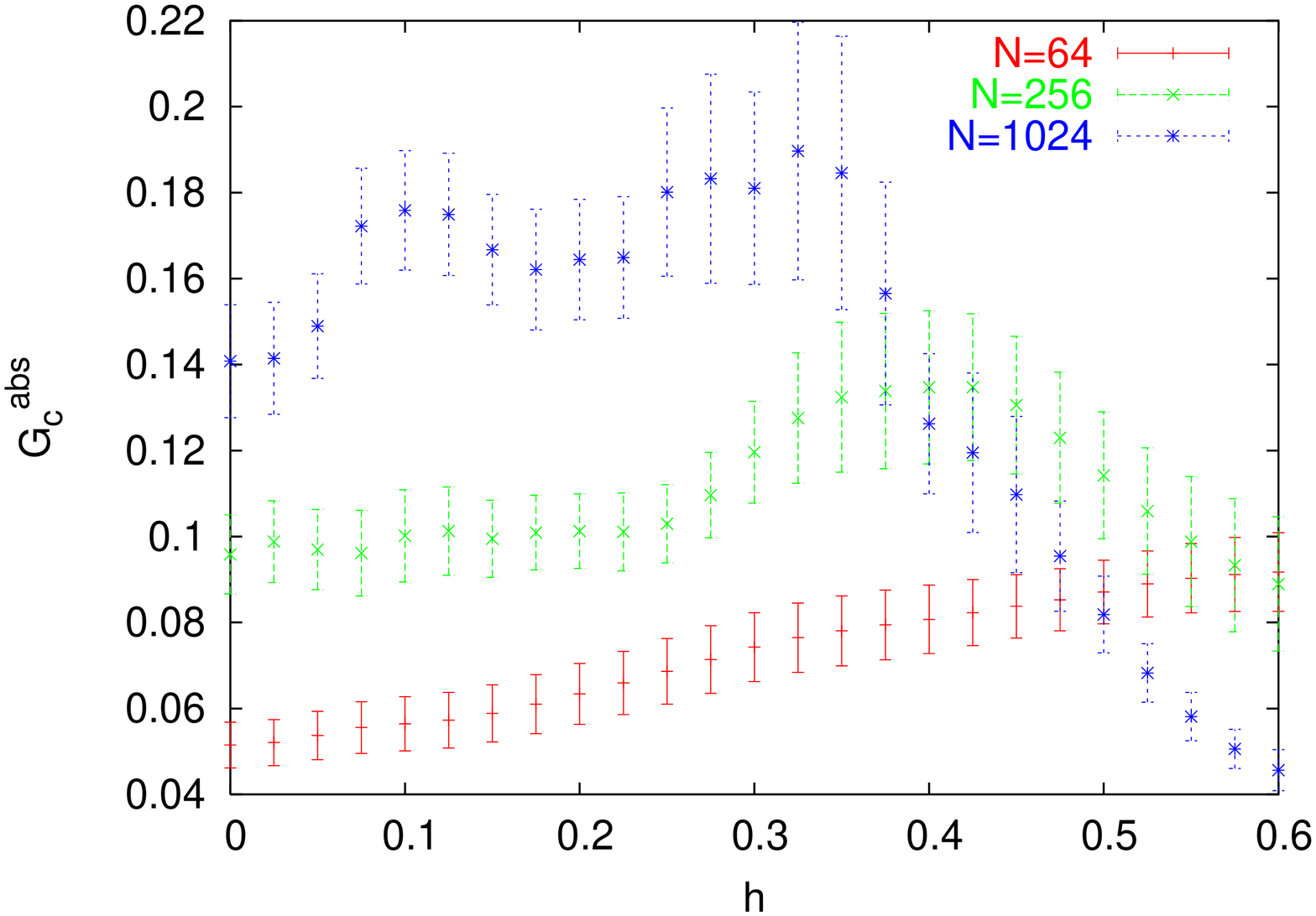,angle=270,width=8cm}
\epsfig{figure=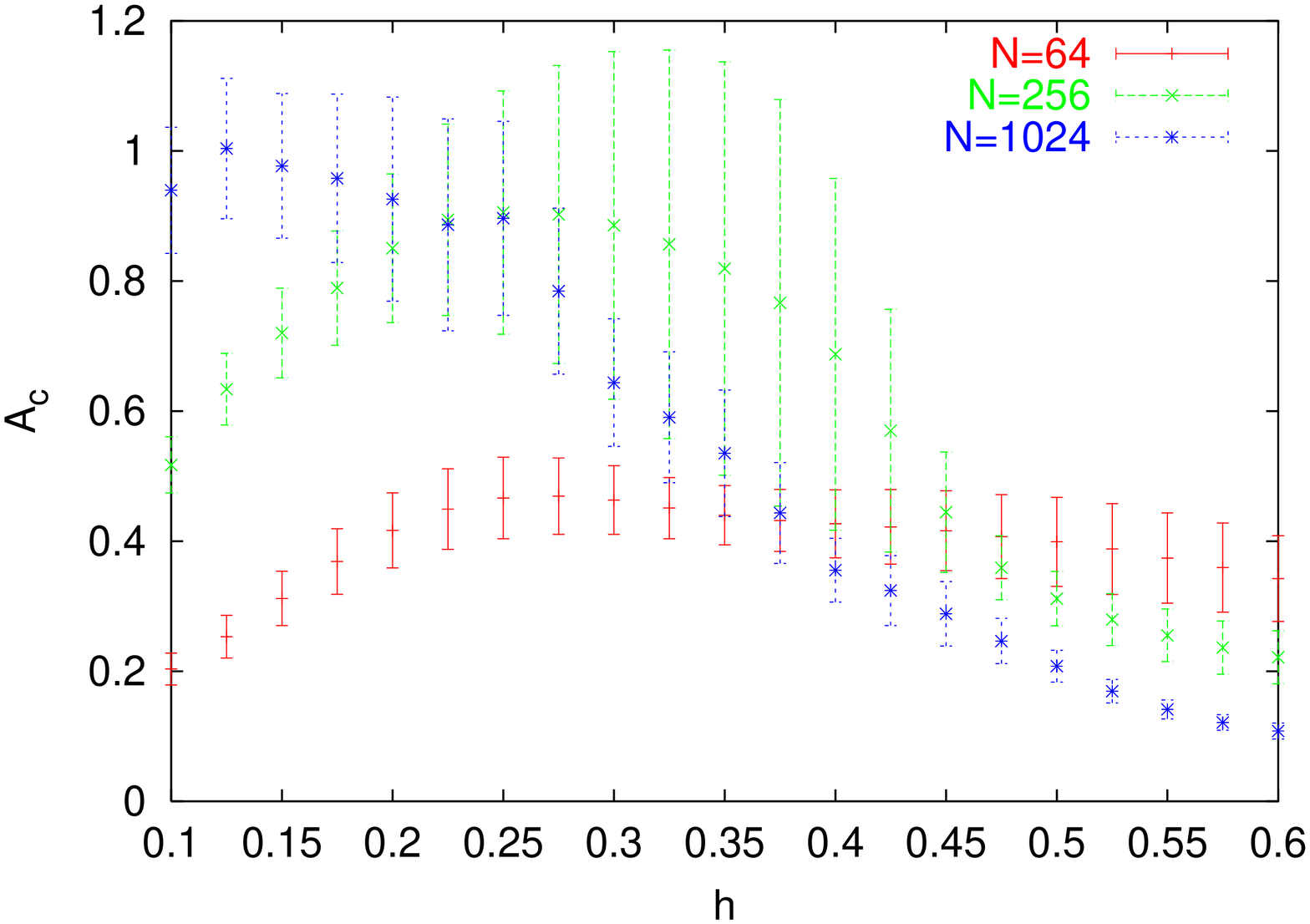,angle=270,width=8cm}
\epsfig{figure=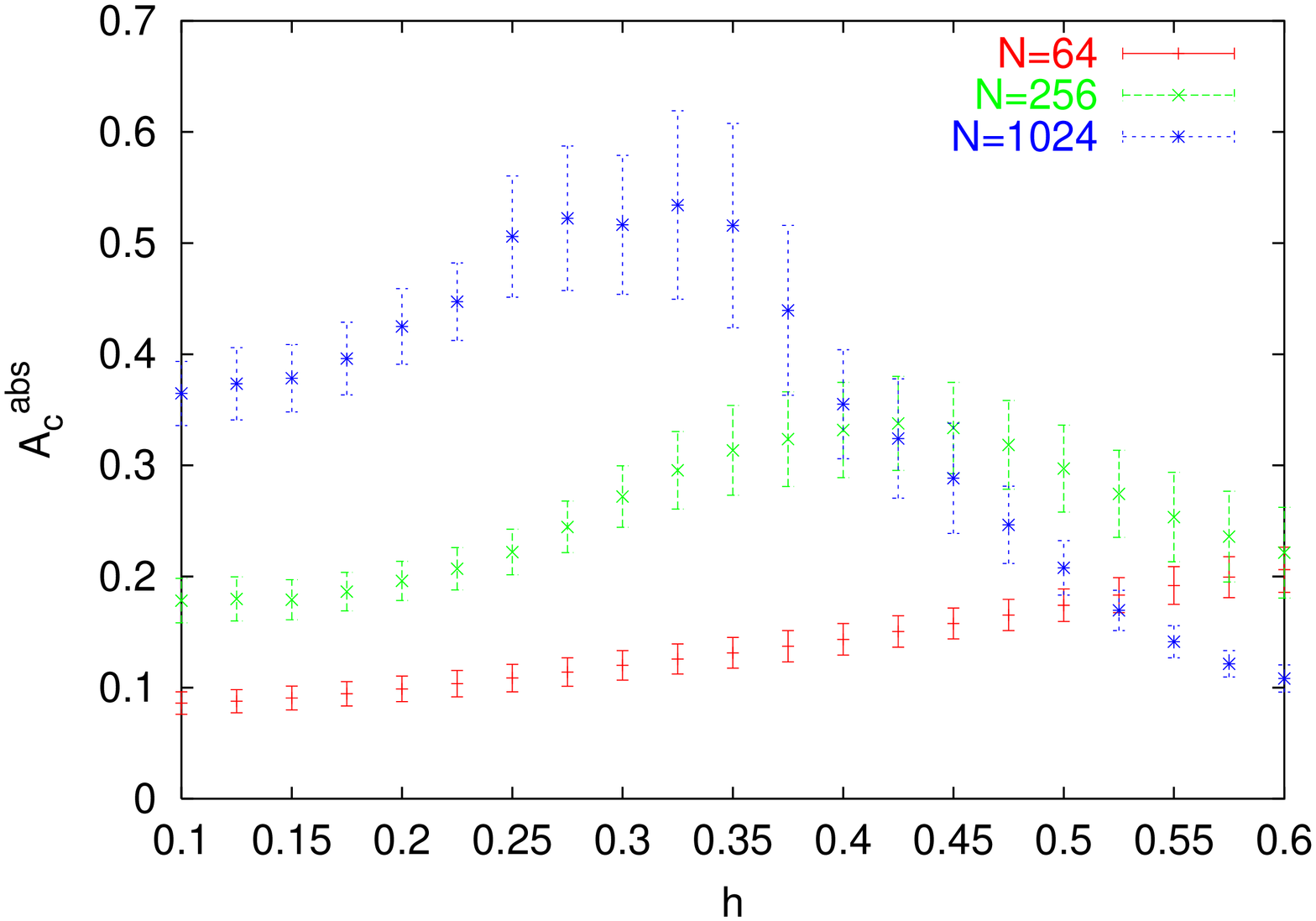,angle=270,width=8cm}
\caption{The behavior of the parameters $G(h,T)$ and
$A(h,T)$ (top), $G_c(h,T)$ and $G_c^{abs}(h,T)$ (center), $A_c(h,T)$
and $A^{abs}_c(h,T)$ (bottom) at $T=0.6$, as a function of the magnetic
field for the different system sizes.}
\end{center}
\end{figure}

\begin{figure}[htbp]
\begin{center}
\leavevmode
\epsfig{figure=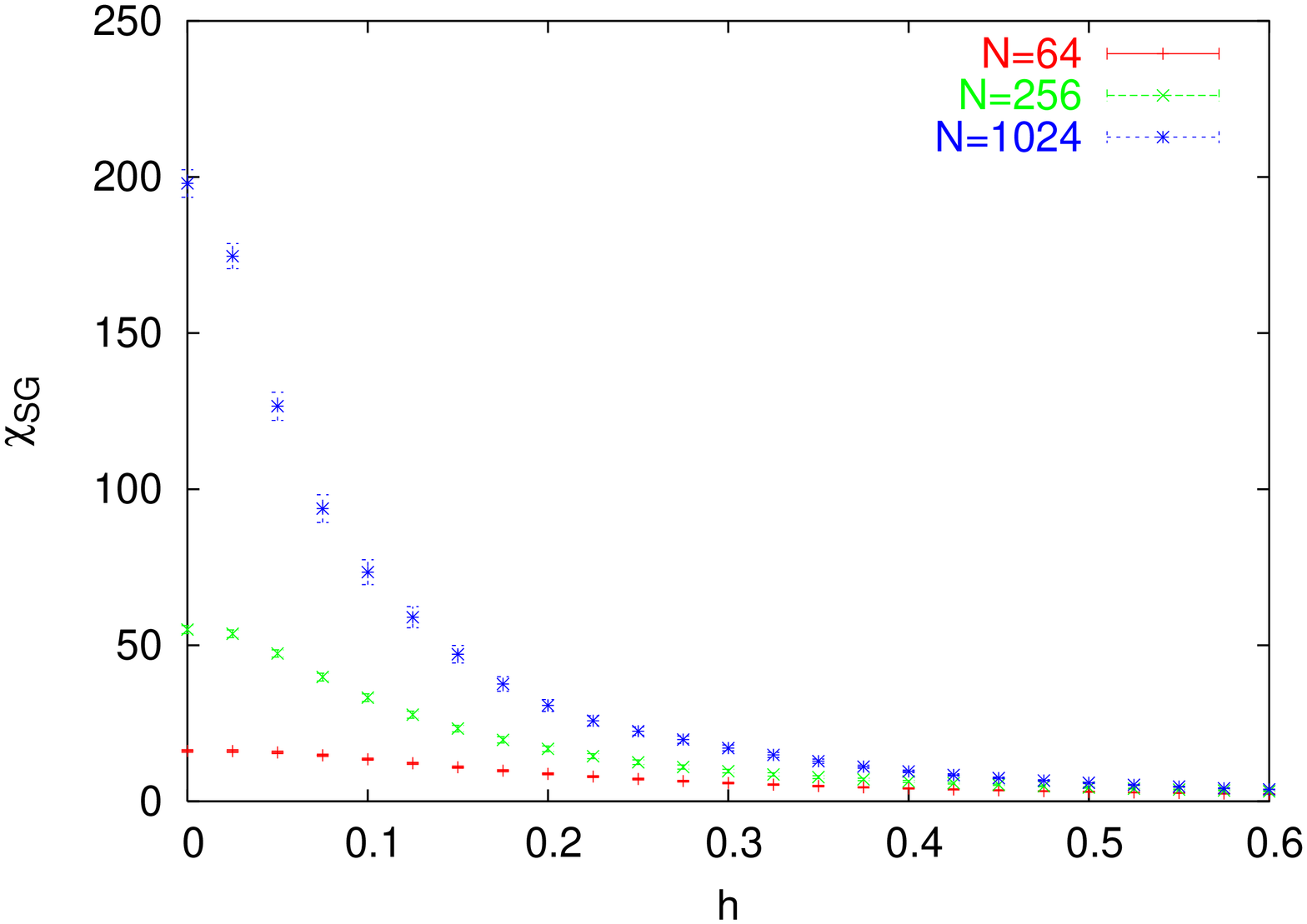,angle=270,width=8cm}
\epsfig{figure=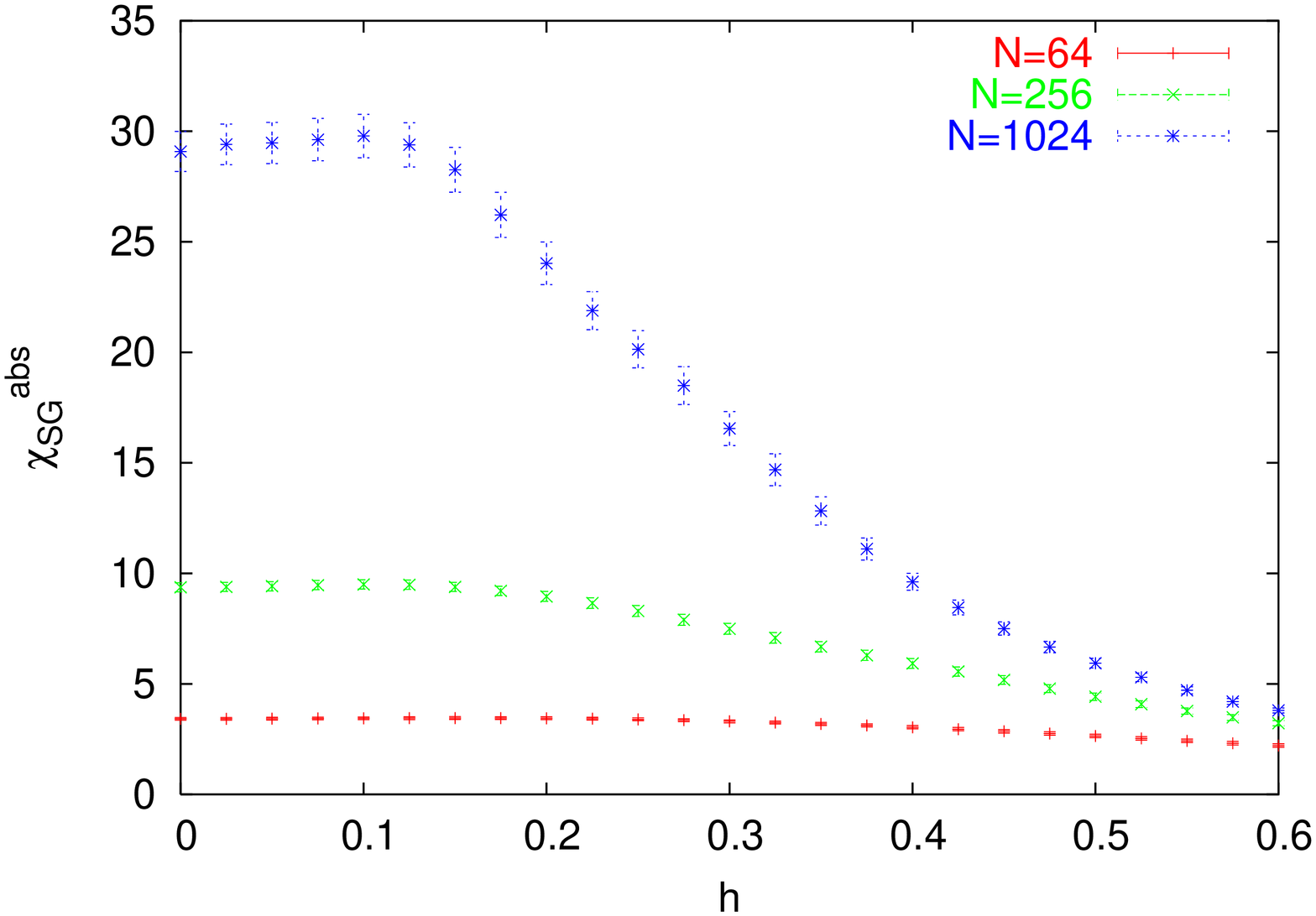,angle=270,width=8cm}
\epsfig{figure=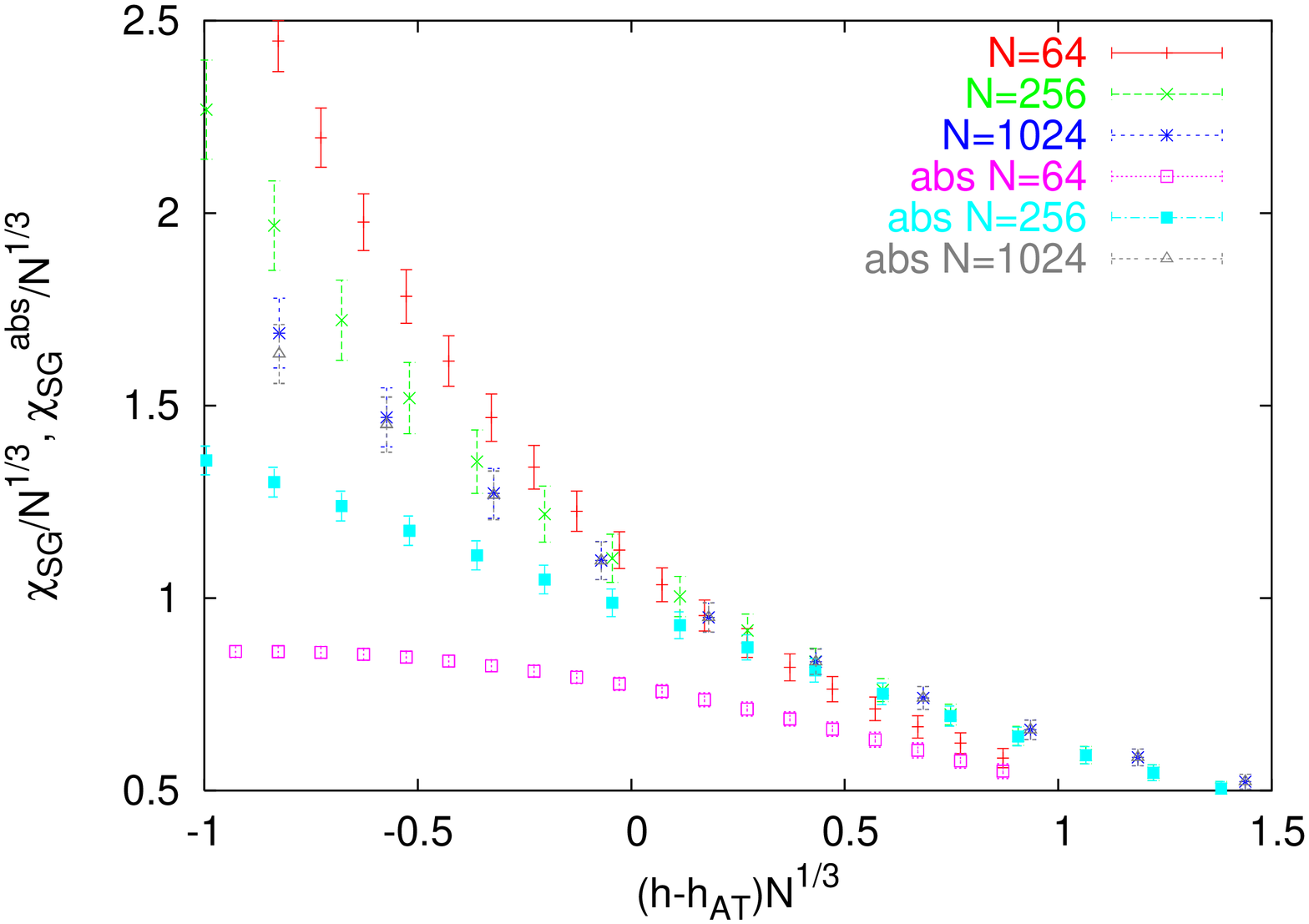,angle=270,width=12cm}
\caption{On the top, the behavior of the spin glass susceptibilities 
$\chi_{SG}(h,T)$ (left) $\chi^{abs}_{SG}(h,T)$ (right).  On the
bottom, scaling plots, i.e.  $\chi_{SG}(h,T)/N^{1/3}$ and
$\chi^{abs}_{SG}(h,T)/N^{1/3}$ plotted as functions of the scaling
variable $(h-h_{AT})N^{1/3}$ at $T=0.6$, for the different system
sizes.}
\end{center}
\end{figure}

An even better evidence for the presence of a phase transition comes from the
behavior of the spin glass susceptibility which is clearly diverging 
(see [Fig. 5]) when entering in the spin glass phase. We see that the
behavior of $\chi_{SG}^{abs}$ (plotted in the same [Fig. 5]) is definitely
different  for the largest size considered $N=1024$ also. Here the 
susceptibility seems to approach a constant in the small field region, 
i.e. for $h \le h_{min}(N)$. Nevertheless $h_{min}(N)$ is clearly 
approaching zero for increasing sizes and also in this case one gets evidence 
for a diverging spin glass susceptibility.

The importance of considering both the usual $\chi_{SG}$ and the
corresponding ``absolute'' quantity $\chi_{SG}^{abs}$ becomes evident
when looking at the scaling plot presented in [Fig. 5]. We find strong
corrections to scaling in the glassy phase, such that it would be hard
to evaluate the correct critical point and exponents from these data.
On the other hand, corrections to $\chi_{SG}$ are in the opposite
direction than  those on $\chi_{SG}^{abs}$, and therefore it is useful
to look at both quantities for understanding the true scaling
behavior. We also note that data for $\chi_{SG}$ and $\chi_{SG}^{abs}$
are nearly coincident in the whole relevant interval (i.e. down to $h
\simeq 0.3$) for $N=1024$, which means that when looking at sizes of
this order or larger one can expect to find not too important
corrections to scaling.

\end{subsection}

\begin{subsection}{The sum rules}
\noindent
Our next aim is to investigate the validity of the stochastic
stability sum rules (\ref{sumrule1}-\ref{sumrules}). Some are supposed to be
non-trivially valid for $h<h_{AT}$, and trivially valid for
$h>h_{AT}$, namely $R^a_{1234}$, $R^a_{1213}$, $R^2_{1234}$ and
$R^2_{1213}$. On the other hand relations $R^b_{1234}$ and $R^b_{1213}$,
that are ratio of moments,
are supposed to be only valid below the AT line. We will also consider
relations $R^{a,abs}_{1234}$ and $R^{b,abs}_{1234}$, obtained from the
probability distribution of the absolute values of the overlap.

We first look at the behavior, as function of the magnetic field, of
the different terms entering the sum-rules, considering the case of
$N=1024$, where finite size corrections are less important.  We see
from [Fig. 6] that
\be
\overline{\langle q_{12}q_{34} \rangle} \simeq  
\overline{\langle | q| \rangle^2} \simeq  
\overline{\langle q_{12}q_{13} \rangle} \simeq 
\overline{\langle q \rangle}^2 \simeq
\overline{\langle q^2 \rangle}
\ee 
for $h\ge0.4$, which is quite reasonable since this is the replica symmetric
region where  $P(q)$ is self-averaging. The tail of  $P(q)$ in the negative 
overlap region has practically disappeared for $N=1024$, so that 
$\langle q \rangle \simeq \langle | q | \rangle$,
and we have moreover (assuming
a Gaussian distribution with variance $\sigma^2$):
\be
\overline{\langle q^2 \rangle}=\sigma^2+q^2_{EA}\simeq q^2_{EA}=
\overline{\langle q \rangle}^2. 
\ee

\begin{figure}[htbp]
\begin{center}
\leavevmode
\epsfig{figure=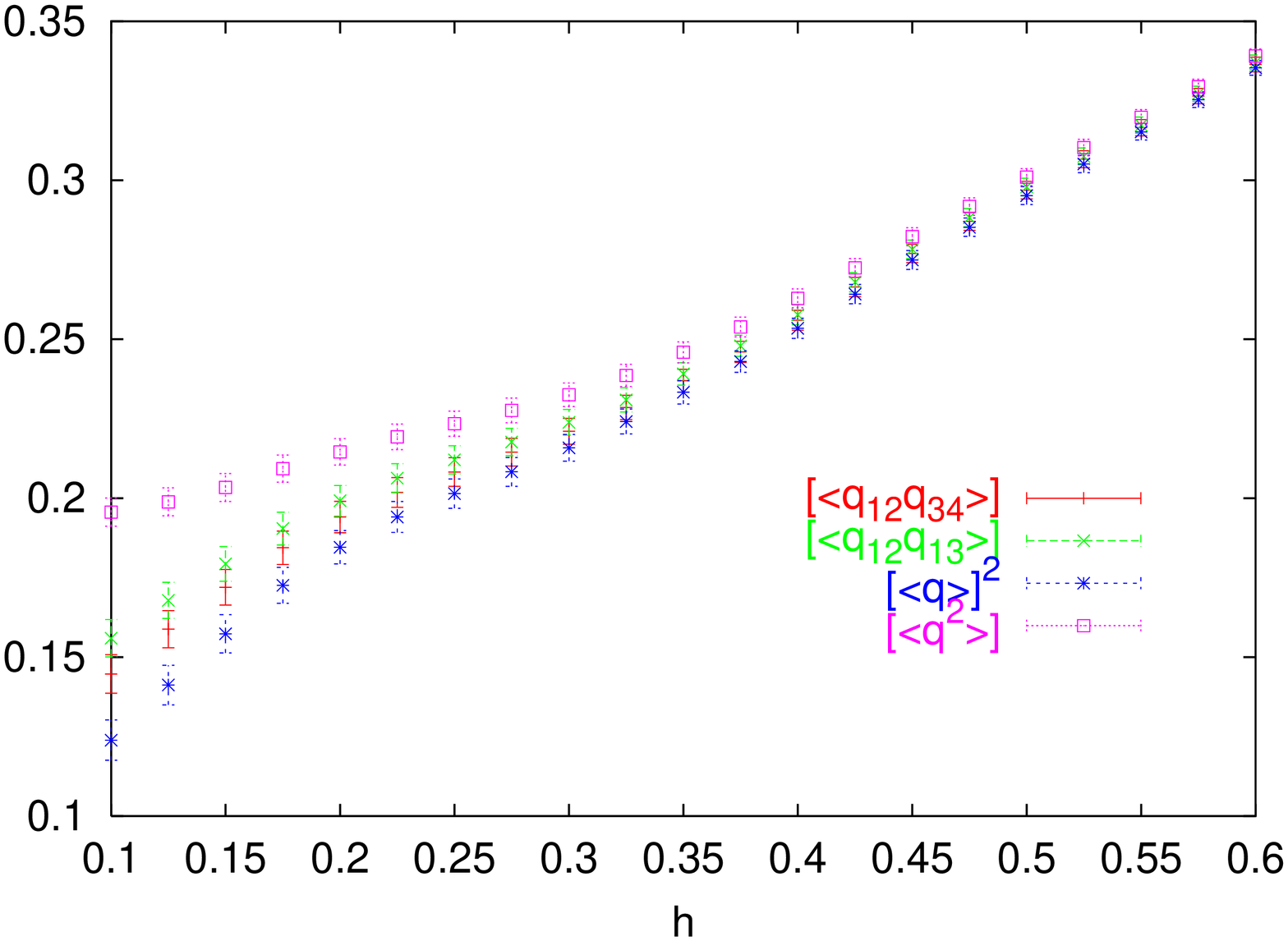,angle=270,width=8cm}
\epsfig{figure=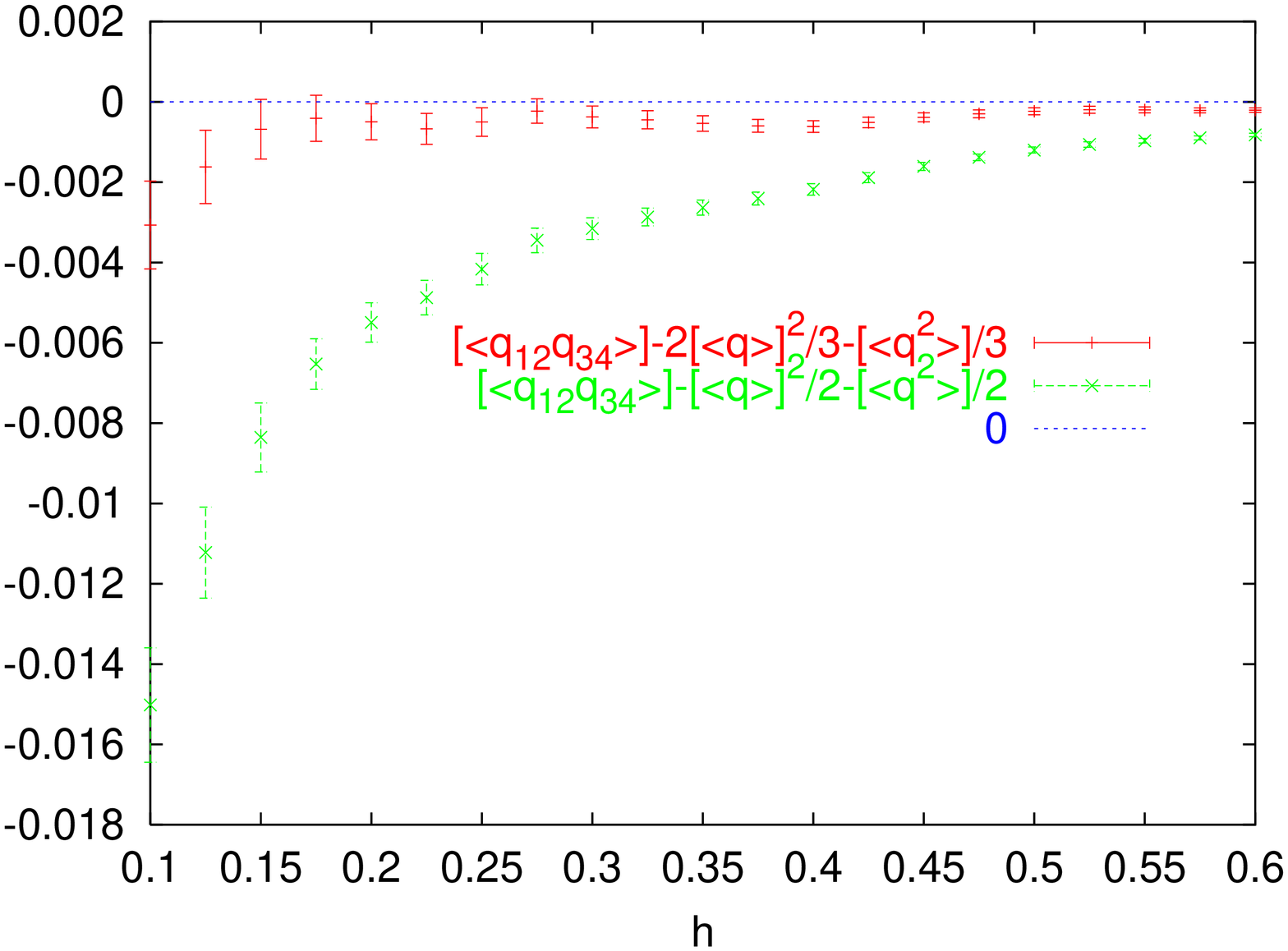,angle=270,width=8cm}
\epsfig{figure=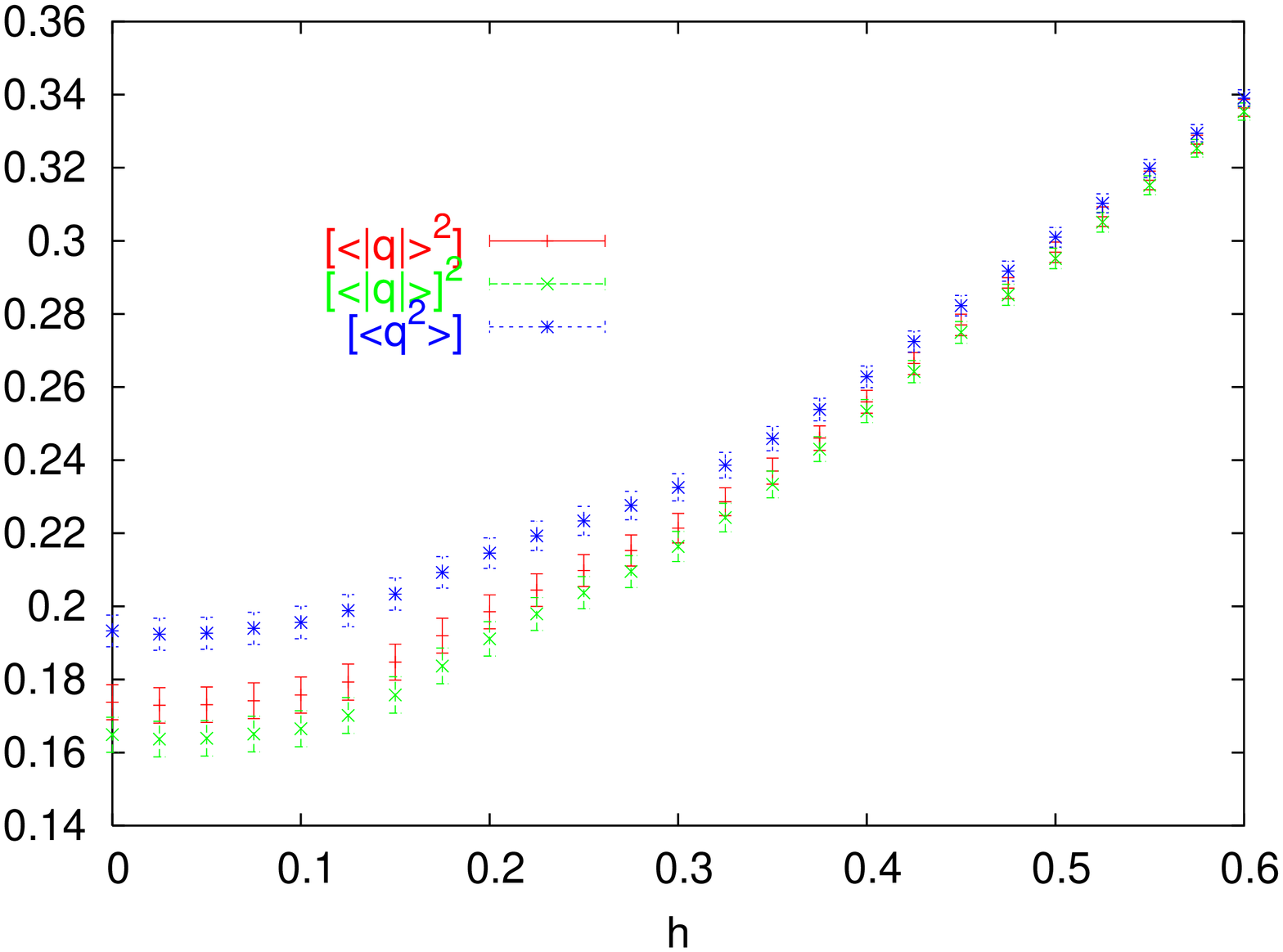,angle=270,width=8cm}
\epsfig{figure=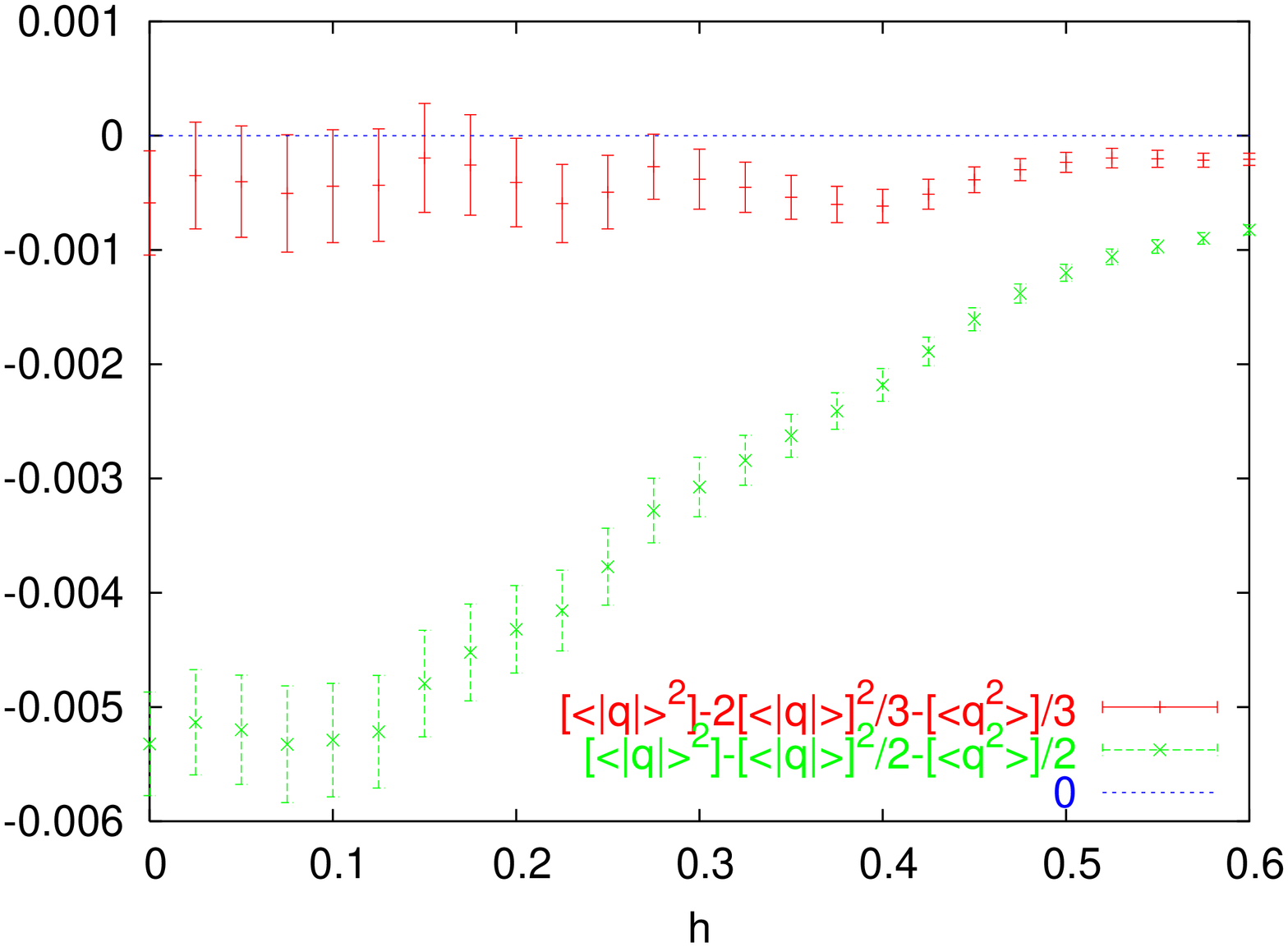,angle=270,width=8cm}
\epsfig{figure=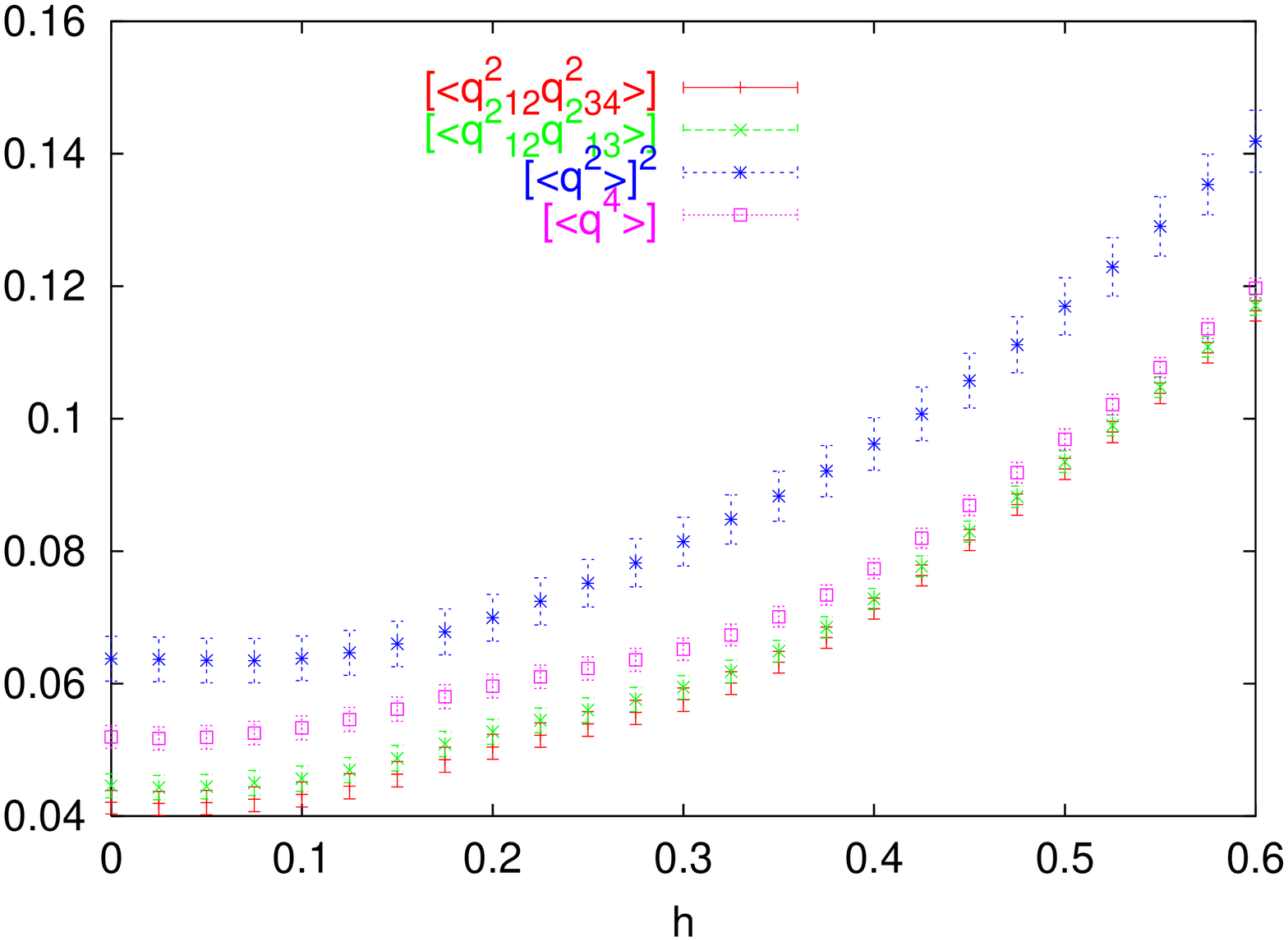,angle=270,width=8cm}
\epsfig{figure=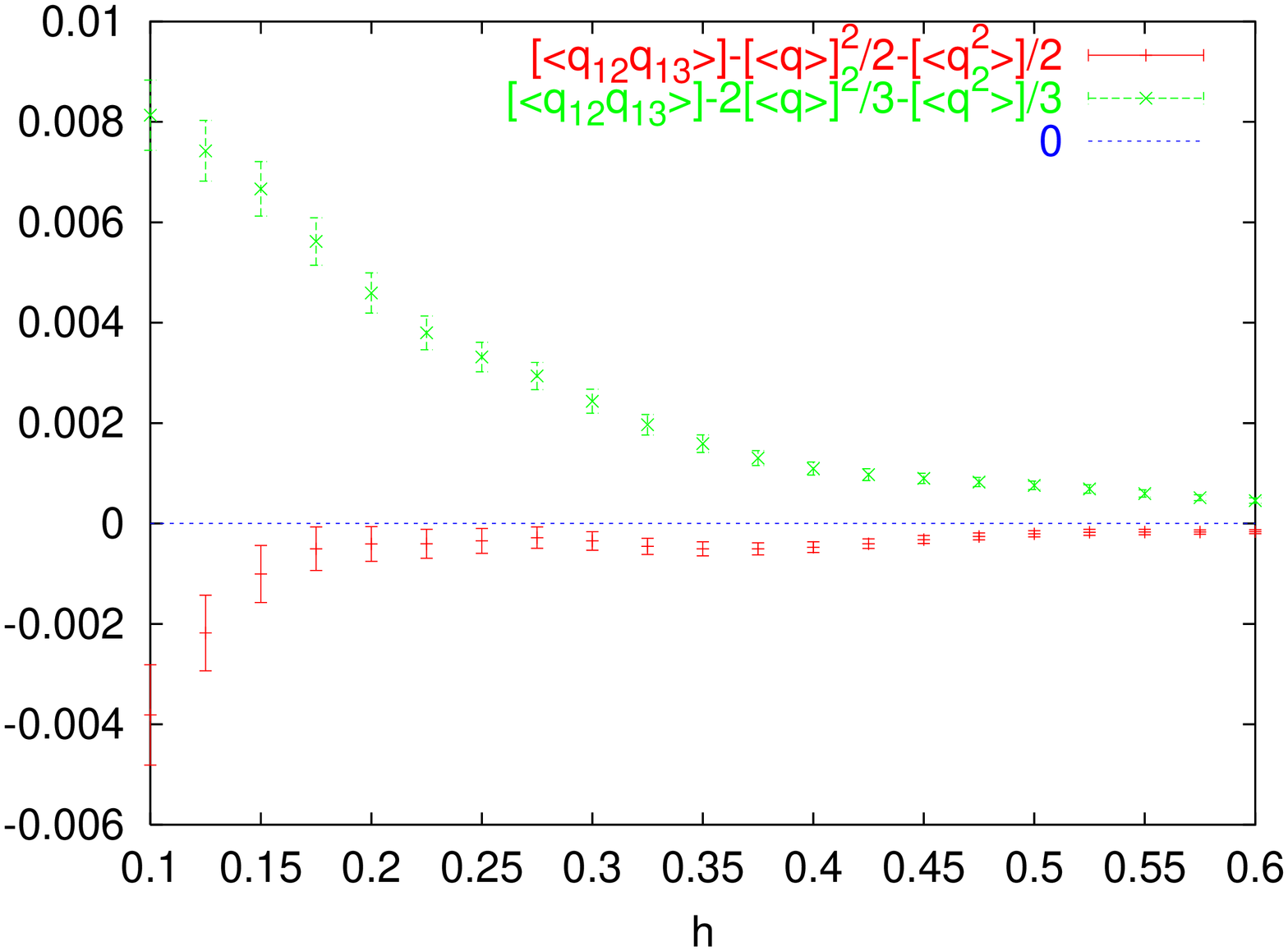,angle=270,width=8cm}
\caption{On the left, we compare the behaviors of 
$\overline{\langle q_{12}q_{34} \rangle}, 
\overline{\langle q_{12}q_{13} \rangle},\overline{\langle q \rangle}^2$ and 
$\overline{\langle q^2 \rangle}$ (top),
$\overline{{\langle |q| \rangle}^2}, 
\overline{\langle |q| \rangle}^2$ and 
$\overline{\langle q^2 \rangle}$ (center), and
$\overline{\langle q^2_{12}q^2_{34} \rangle}, 
\overline{\langle q^2_{12}q^2_{13} \rangle},\overline{\langle q^2 \rangle}^2$ 
and $\overline{\langle q^4 \rangle}$ (bottom) as function of the
magnetic field for $N=1024$. On the right we plot (for $N=1024$ again) relations
$R^{a}_{1234}$ (top), $R^{a,abs}_{1234}$ (center) 
and $R^{a}_{1213}$ (bottom), that are well satisfied,
together with the modified ones $T_{1234}$, $T^{abs}_{1234}$ and 
$T_{1213}$ respectively, that are not in the SG phase (see text).}
\end{center}
\end{figure}

\begin{figure}[htbp]
\begin{center}
\leavevmode
\epsfig{figure=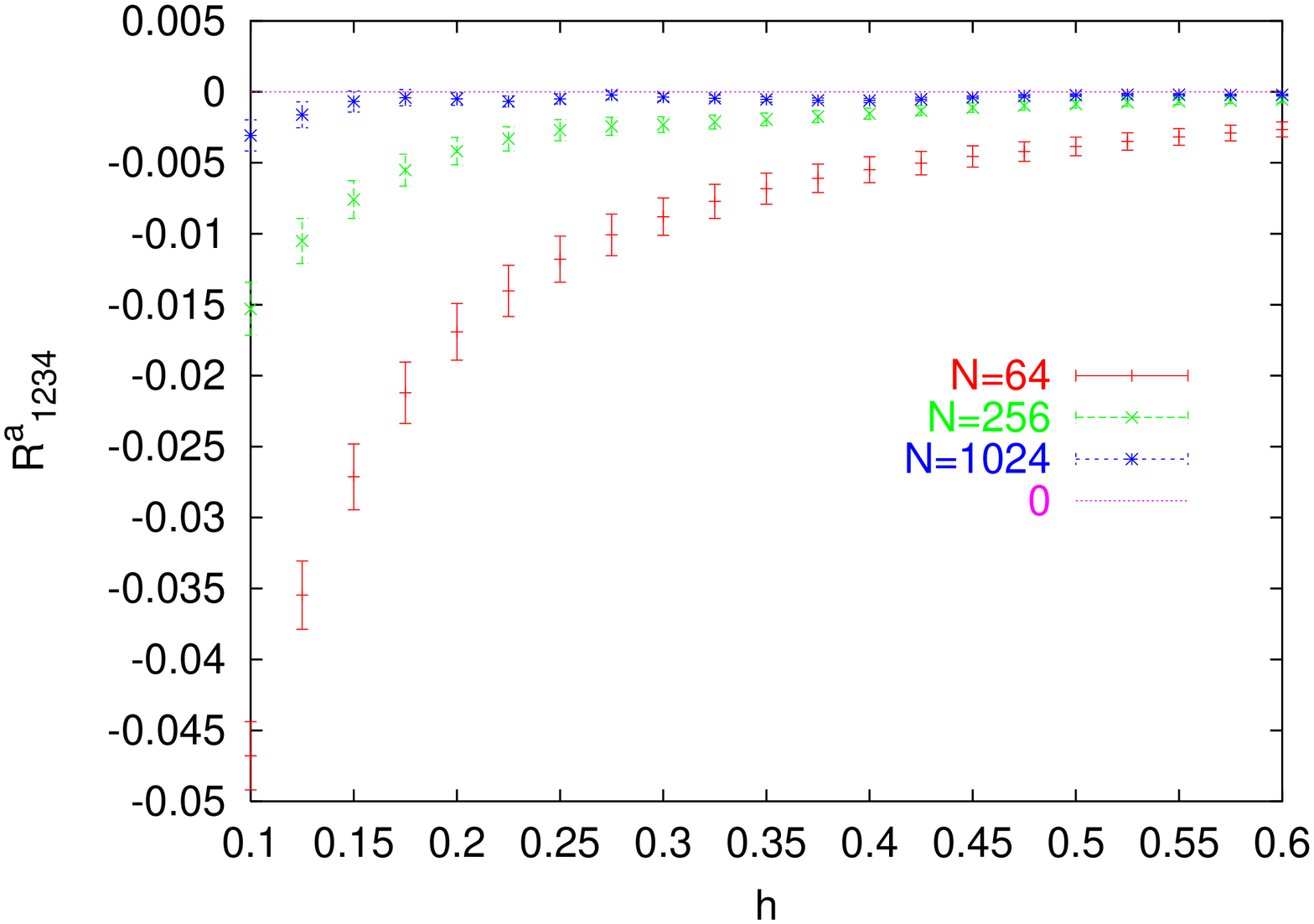,angle=270,width=8cm}
\epsfig{figure=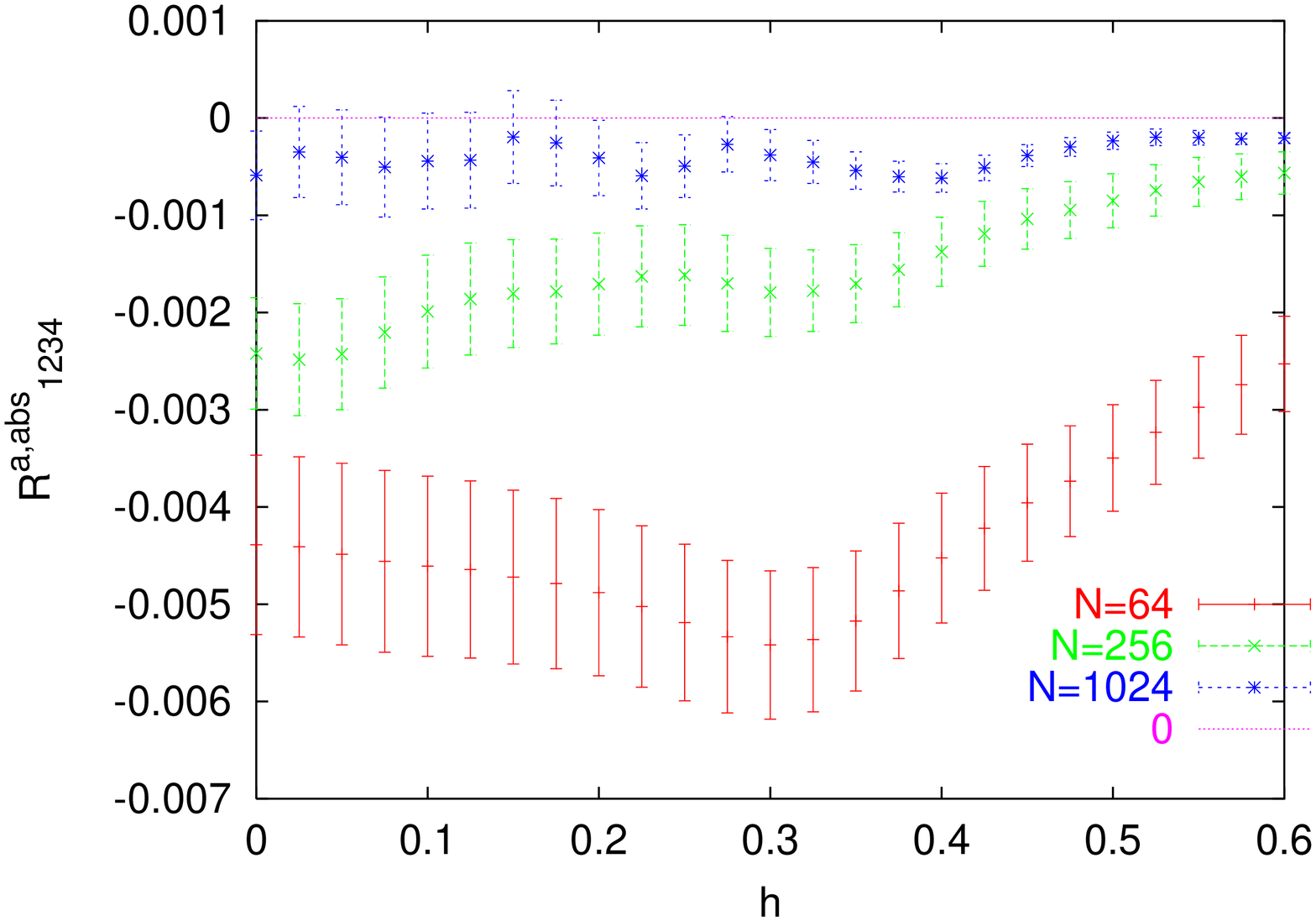,angle=270,width=8cm}
\epsfig{figure=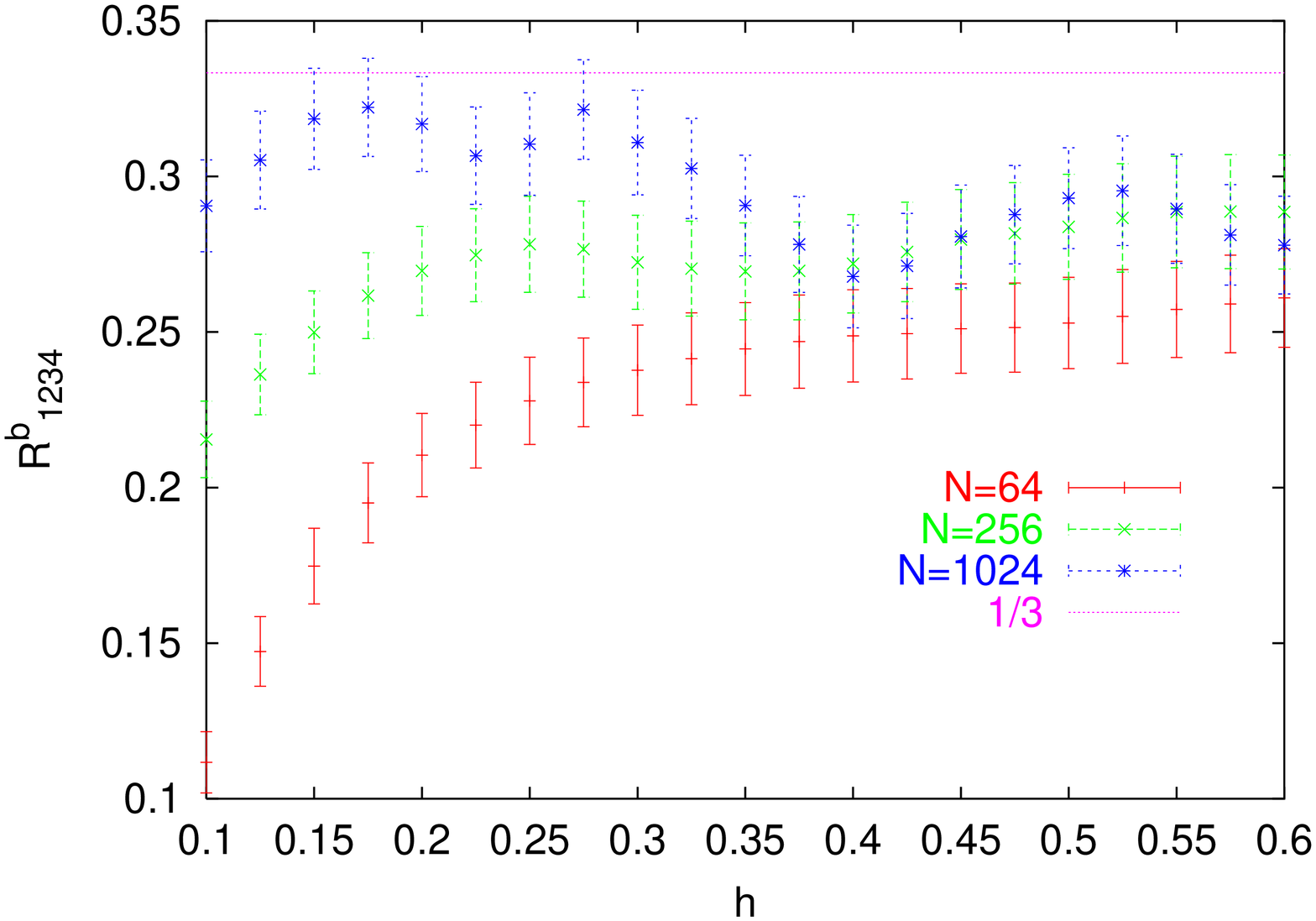,angle=270,width=8cm}
\epsfig{figure=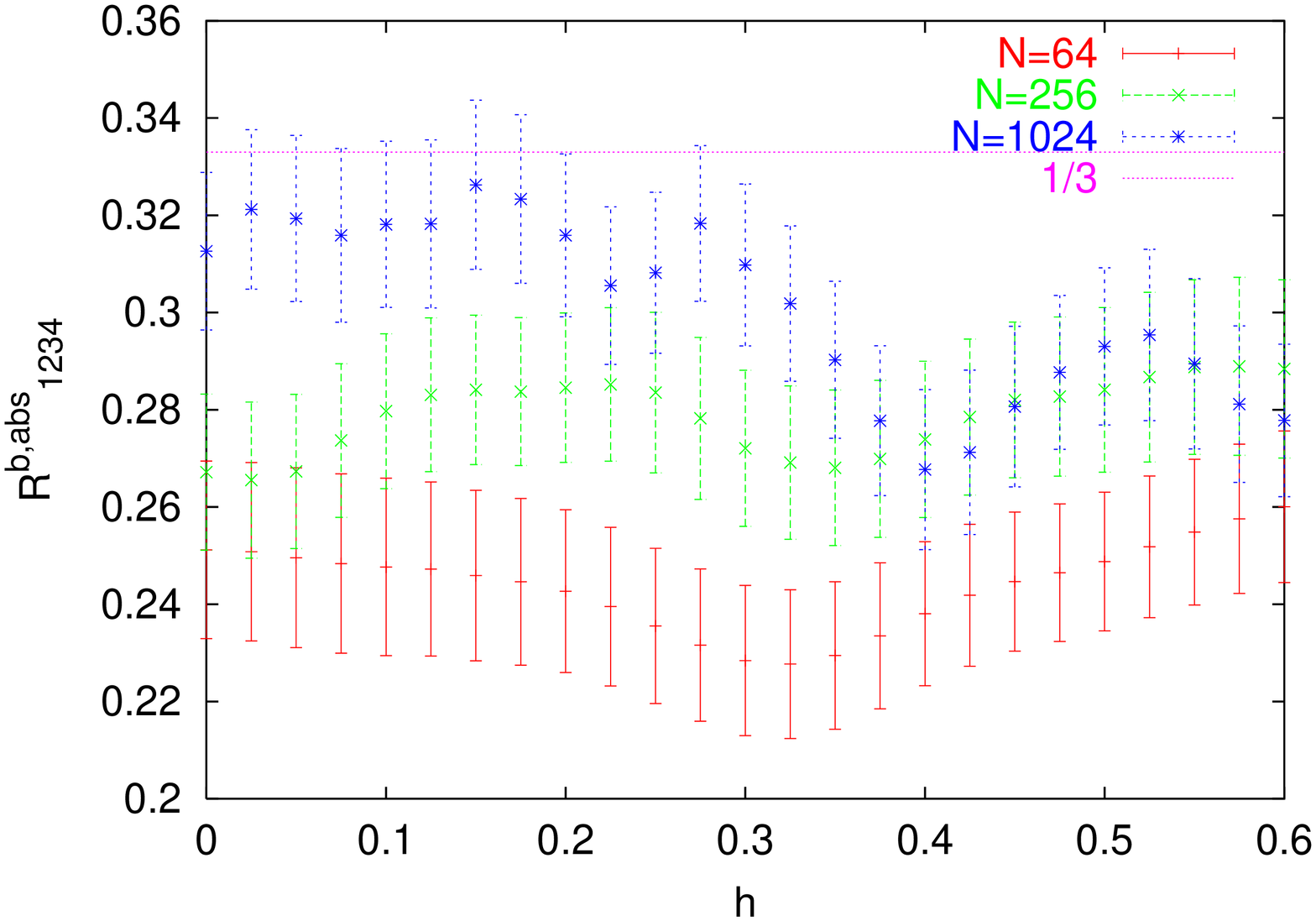,angle=270,width=8cm}
\epsfig{figure=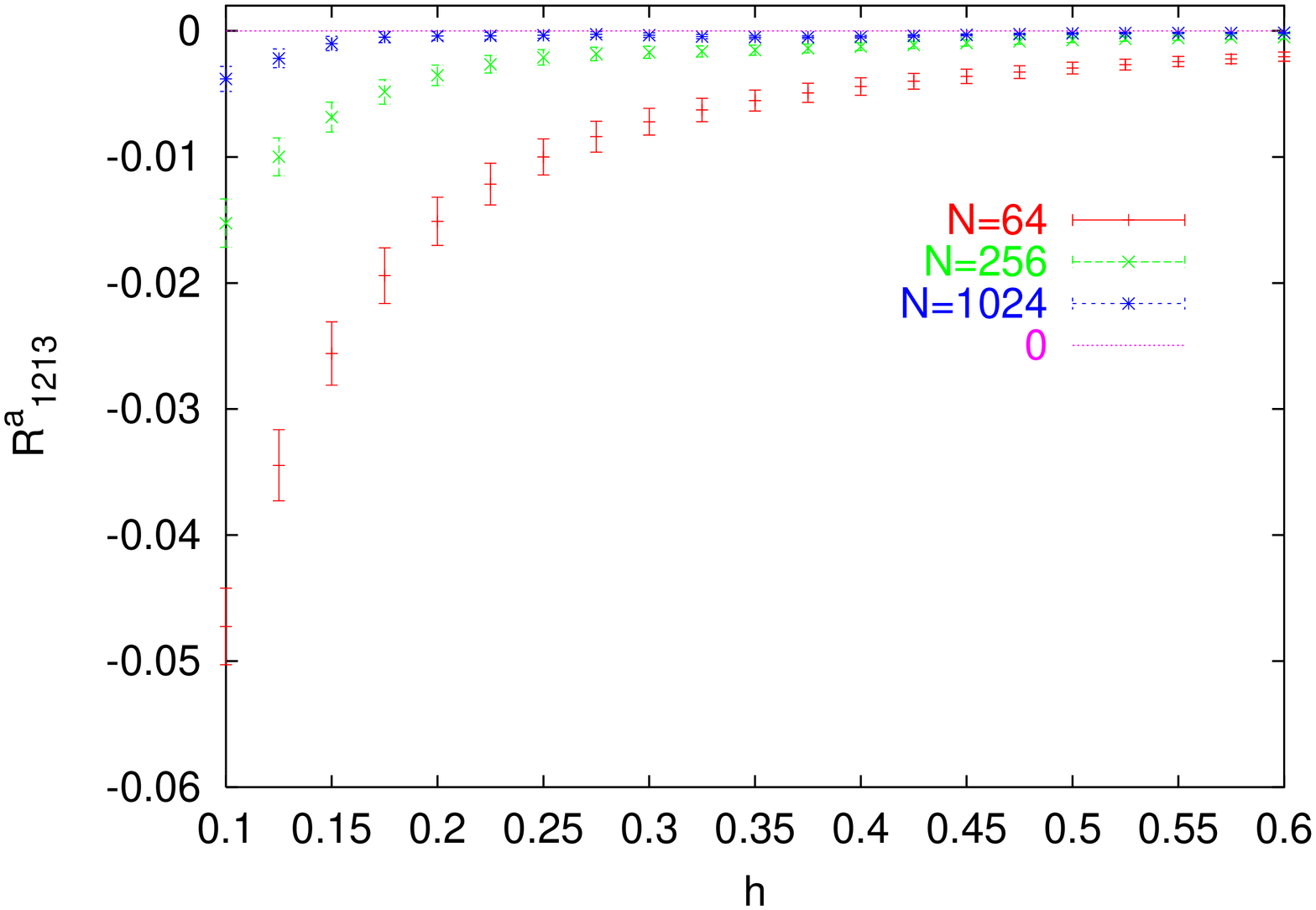,angle=270,width=8cm}
\epsfig{figure=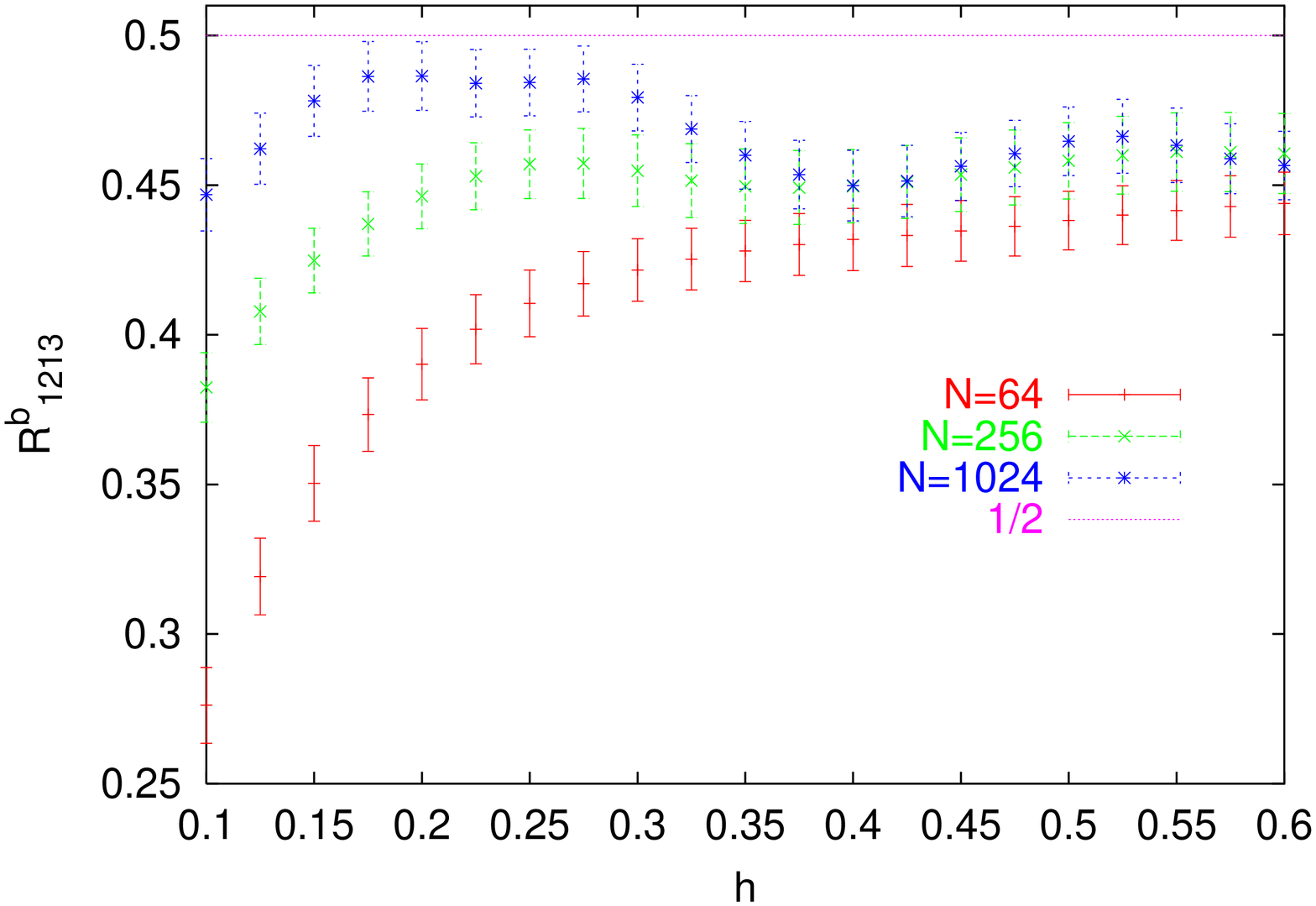,angle=270,width=8cm}
\caption{The behavior of $R^{a}_{1234}$ (top,left), $R^{a,abs}_{1234}$ (top,right),
$R^{b}_{1234}$ (center,left), $R^{b,abs}_{1234}$ (center,right),
$R^{a}_{1213}$ (bottom, left) and $R^{b}_{1213}$ (bottom, right) as
function of the magnetic field for the different considered system
sizes.}
\end{center}
\end{figure}

\begin{figure}[htbp]
\begin{center}
\leavevmode
\epsfig{figure=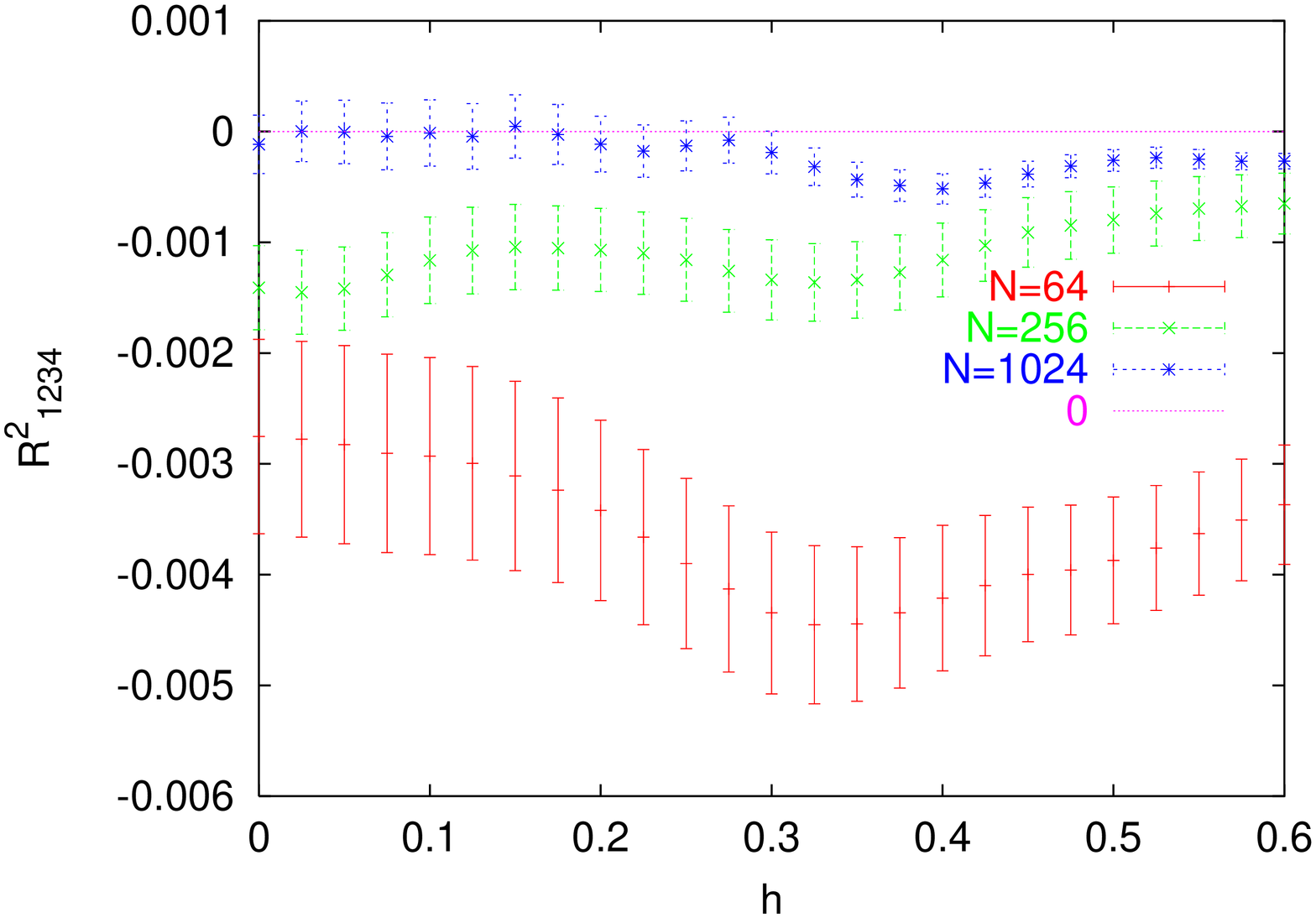,angle=270,width=8cm}
\epsfig{figure=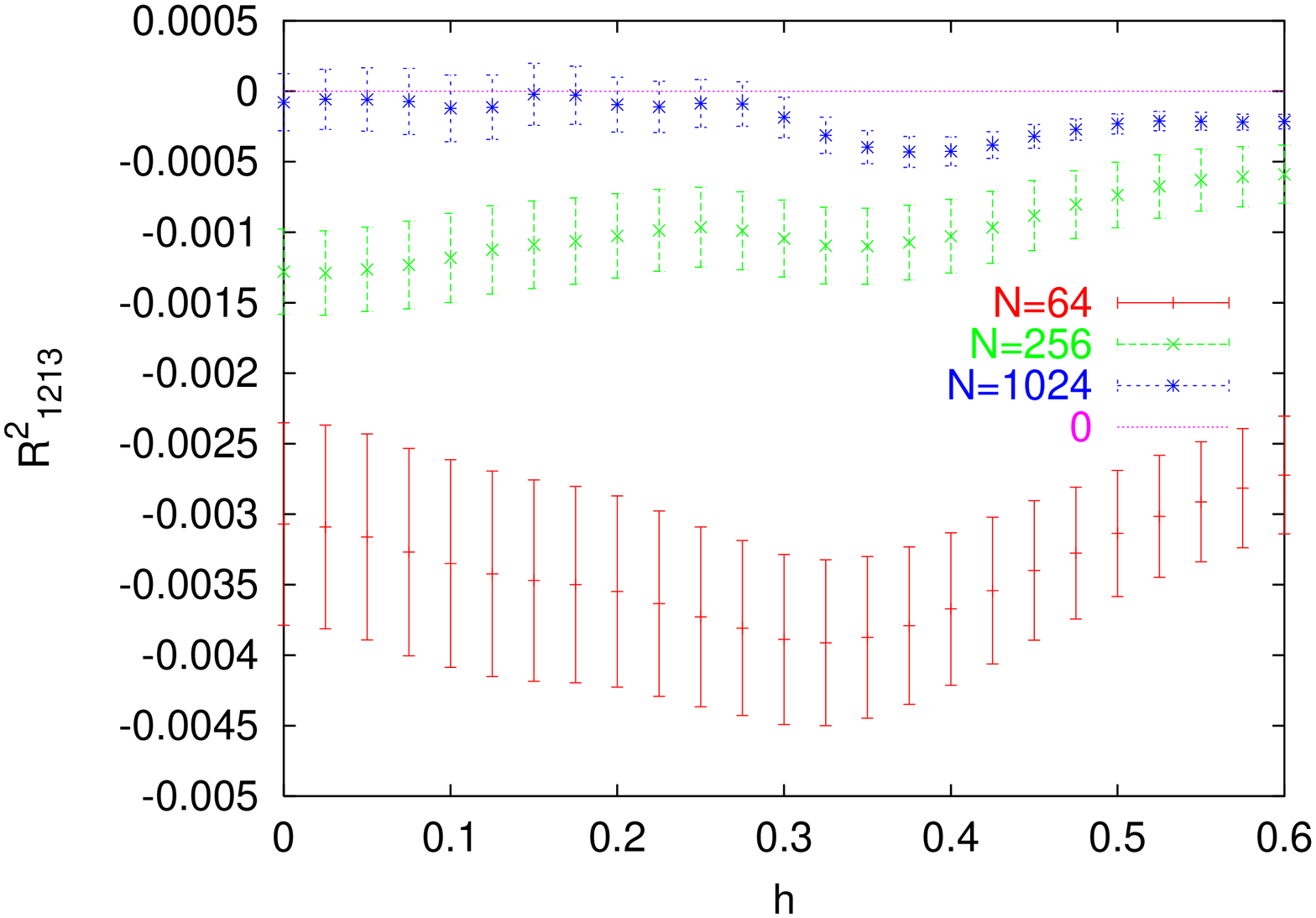,angle=270,width=8cm}
\caption{The behavior of $R^2_{1234}$ (left) and $R^2_{1213}$ (right) as
a function of the magnetic field for the different  system sizes.}
\end{center}
\end{figure}

On the other hand, these quantities are different when entering into
the glassy phase, though the differences appear small.

We plot in the same [Fig. 6] relations $R^a_{1234}$,
$R^{a,abs}_{1234}$ and $R^a_{1213}$. In order to show that they are
non-trivially verified we plot also, following \cite{MaPaRiRuZu}, the relations
\begin{eqnarray}
T_{1234}(h,T)&=&\overline{\langle q_{12}q_{34} \rangle}-{1\over 2}
\overline{\langle q \rangle}^2-
{1\over 2} \overline{\langle q^2 \rangle} =0 \\
T^{abs}_{1234}(h,T)&=&\overline{\langle | q|  \rangle^2}-{1\over 2}
\overline{\langle | q | \rangle}^2-
{1\over 2} \overline{\langle q^2 \rangle}  =0 \\
T_{1213}(h,T)&=&\overline{\langle q_{12}q_{13} \rangle}-{2\over 3}\overline{\langle q \rangle}^2-
{1\over 3} \overline{\langle q^2 \rangle} =0. 
\end{eqnarray}
which should also be verified if $R^a_{1234}$, $R^{a,abs}_{1234}$ and 
$R^a_{1213}$ where trivial. They are clearly not verified in the spin-glass 
phase and accordingly $R^a_{1234}$ and $R^a_{1213}$ are non-trivial in this  phase.
We moreover note that to look at the probability distribution of the absolute value 
overlap is very useful also in this case, since as we already pointed out 
these relations are derived  assuming an infinitesimal magnetic field 
which breaks the global symmetry for inversion of all the spins. As a 
matter of fact, relations $R^a_{1234}$ and $R^{a}_{1213}$ are no more verified
as soon as $h\siml 0.15$, where the tail of  $P(q)$  in the
negative overlap region becomes important also for $N=1024$ (we do not
present  data for $h\le 0.1$). On the other hand, relation 
$R^{a,abs}_{1234}$ appears very well satisfied within the errors down
to $h=0$.

In [Fig. 7] we present our data for relations $R^{a}_{1234}$,
$R^{a,abs}_{1234}$ and $R^{a}_{1213}$ with a finer vertical scale than
in [Fig. 6], together with data for $R^{b}_{1234}$, $R^{b,abs}_{1234}$ and
$R^{b}_{1213}$.  The situation is quite clear: for $N=1024$ relations
$R^{b}_{1234}$, $R^{b, abs}_{1234}$, $R^{b}_{1213}$ are not verified
above the AT line, whereas they are satisfied within the statistical
error below (up to crossover effects for small $h$'s for the non
``absolute'' quantities) .

In any event, the change of behavior in the sum rules when going from
the $h\le 0.4$ region to the other side of the AT line is small. This
is trivial in the case of relations $R^{a}_{1234}$, $R^{a,abs}_{1234}$
and $R^{a}_{1213}$ since all terms become very similar (see [Fig. 6])
as we already discussed. 
This can be understood \cite{Private} for 
$R^{b}_{1234}$, $R^{b,abs}_{1234}$ and
$R^b_{1213}$ (i.e. the ratios $\chi^{1234}_{SG}/\chi_{SG}=
\chi^{abs,1234}_{SG}/\chi_{abs,SG}=1/3$ and $\chi^{1213}_{SG}/\chi_{SG}=1/2$ 
respectively), using the results of  \cite{PiDoTe}. This
$R^b$'s can be calculated from the masses $r_R$, $r_L$ and
$(r_L-r_A)/n$ computed in this paper. Rather surprisingly, the ratio condition
becomes true again in the high field limit, and the $R^b$'s gain back
their $1/2$ and $1/3$ values.  This means that these $R^b$'s have only a
very slight variation in the RS phase, with probably a minimum, and
they are continuous at the AT-line.
We note that their behavior is very
similar to the behavior of $G$, further confirming that appropriate
parameters for getting evidence for the transition are the ones which
involve connected quantities, such as $G_c$ and $A_c$.

In [Fig. 6] we also present the behavior of the different terms
entering the relations $R^2_{1234}$ and $R^2_{1213}$. 
$\overline{ \langle q^2 \rangle^2}$ is definitely different
from the other terms, and quite surprising remains clearly different also 
on the other side of the AT line, which is to be
interpreted as a reminiscence of non-self-averageness due to
finite size effects.

Finally we plot in [Fig. 8] relations $R^2_{1234}$ and $R^2_{1213}$. Here 
finite size effects are less important because these relations are valid
also in the $h\rightarrow 0$ limit. Nevertheless  these quantities are 
compatible with zero within our statistics for $h\siml 0.3$ only for $N=1024$.
Also in this case we find only a small difference between the $h<0.4$ behavior
and the one outside the glassy phase. From this point of view it should be
recalled that $\overline{\langle q^2 \rangle^2}$ is definitely different
from the other terms in the whole $h$ range, which means that data for 
$h>0.4$ 
are far from being in the asymptotic self-averaging regime in which these sum 
rules should be trivially satisfied. As a last remark, we note that from 
relation $R^2_{1234}$ immediately follows the expected behavior of the
parameter $G$ and that the small differences we observe here between the 
behaviors inside and outside the glassy phase do indeed reflect  the
fact that $G$ is not an appropriate observable to look at for obtaining
evidence of the transition.
\end{subsection}

\begin{subsection}{On  $P(q)$ for a large size}
Our data for $P(q)$ at $h=0.3$ for a system of $3200$ spins can be
found in [Fig 9] for $T=0.4$, $0.5$ and $0.6$. The corresponding
values of $q_{EA}$ are $0.759$, $0.640$ and $0.505$ respectively.  We
have been very careful in checking that thermalization is achieved
for all values of the temperature. It is clear from the figure why the
asymptotic behavior of this distribution has escaped observation up to
now. At $T=0.6$ only a single peak (corresponding to $q_{EA}$) is
visible, with substantial asymmetry (the distribution is wider in the
low $q$ side). The asymmetry is stronger for $T=0.5$, but there is
still no sign of the low $q$ peak\footnote{This result disagree with
\cite{PiRi}, where a low $q$ peak is found, using the 
Metropolis algorithm with $100000$ sweeps for equilibrium and only
$20$ disorder samples.}.  Only for $T=0.4$ does one see the expected
continuum on the left of the self-overlap peak, with some indication
of the low $q$ peak at a location in agreement with the
value \cite{CrRi} $q_{min}\simeq 0.44$. It should be noted that the
peak corresponding to $q_{min}$ is predicted to be  broader than the
one corresponding to $q_{EA}$ \cite{FrPaVi}. This explains why we do not 
observe this minimum overlap peak.

\begin{figure}
\begin{center}
\leavevmode
\epsfig{figure=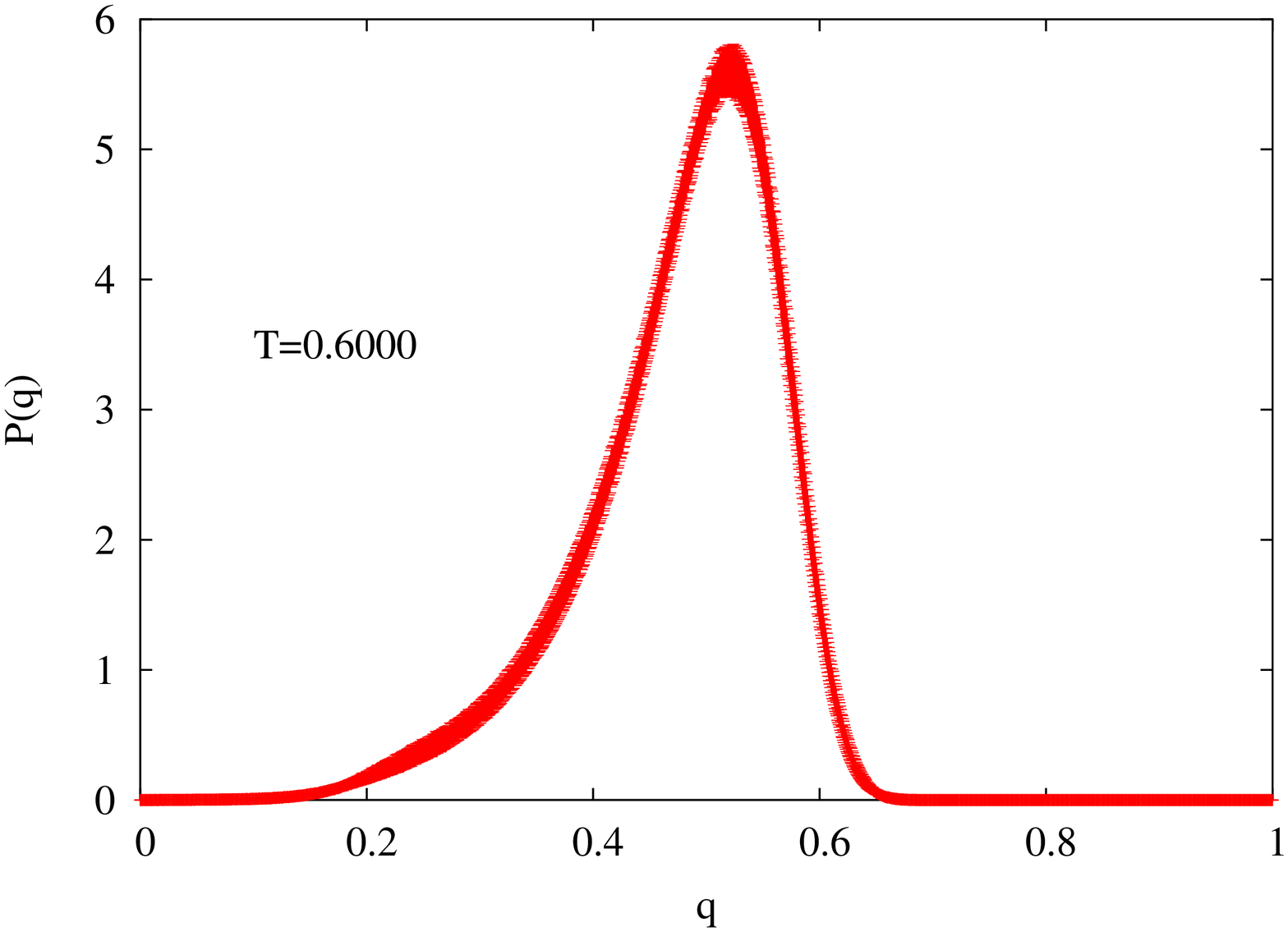,angle=270,width=8cm}
\epsfig{figure=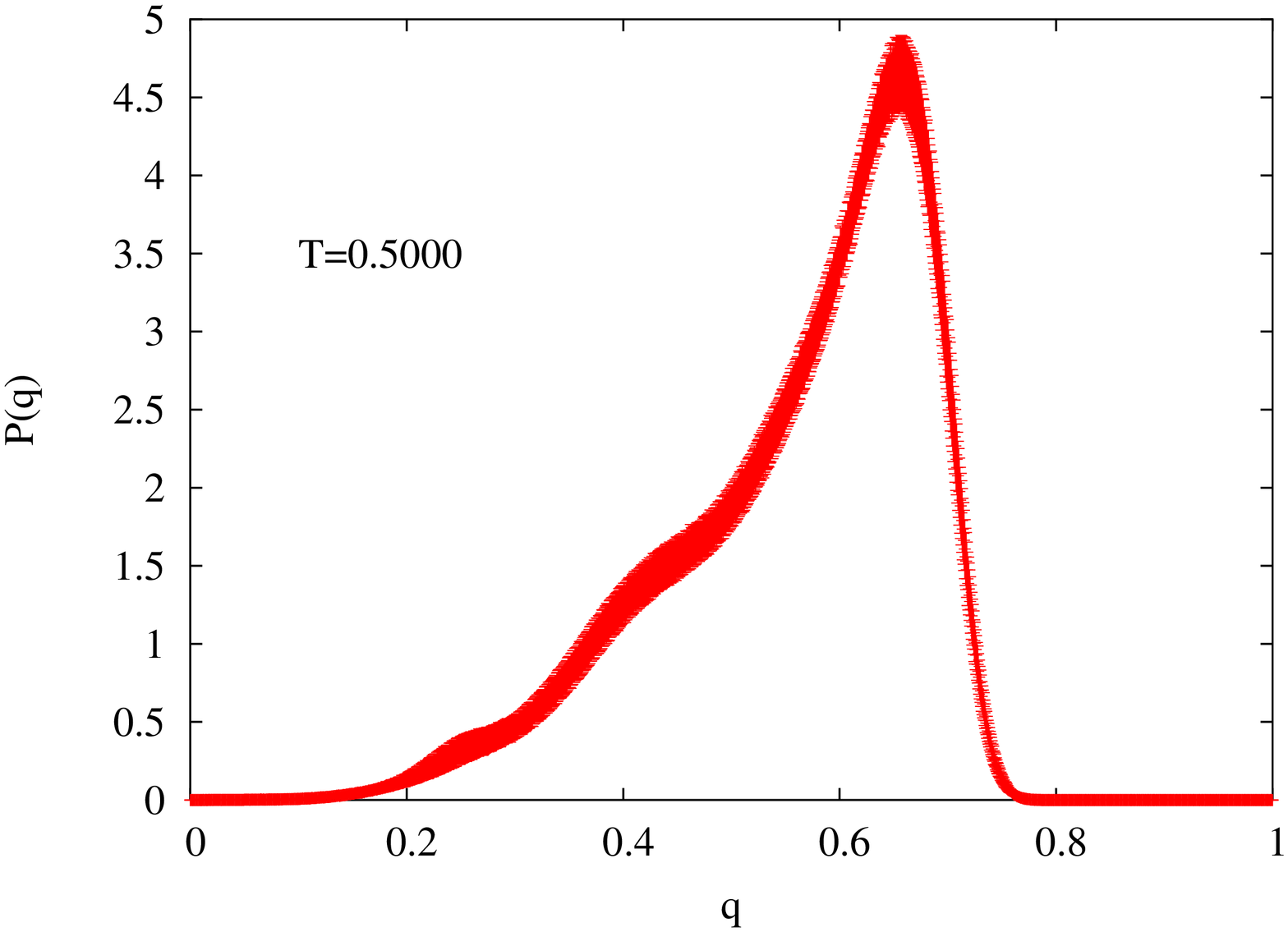,angle=270,width=8cm}
\epsfig{figure=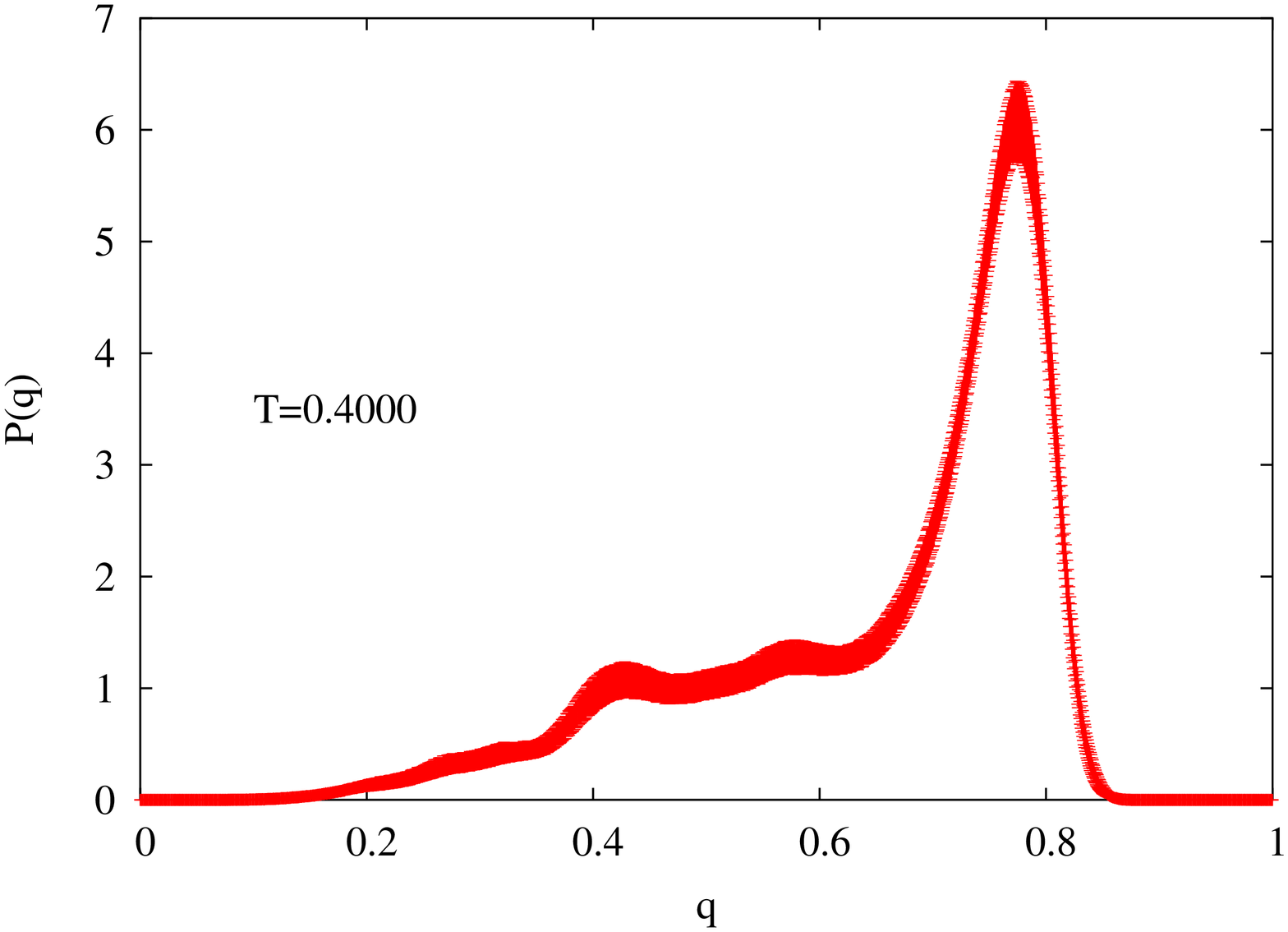,angle=270,width=8cm}
\caption{The behavior of the probability distribution of the overlap
$P(q)$ with $h=0.3$ and temperatures $T=0.4$, 0.5 and 0.6 respectively,
for the large size $N=3200$.}
\end{center}
\end{figure}

\end{subsection}
\end{section}

\begin{section}{Conclusions}
\noindent
We performed numerical simulations of the SK model in a magnetic field
at temperature $T=0.6$, both in the glassy phase and above the AT
line. We used a modified version of the PT algorithm in which the system is
allowed to move between a chosen set of magnetic field values, an algorithm 
well suited  for our purpose.

We measured quantities such as the magnetic susceptibility, which turns out
to be in agreement with the predicted analytical behavior of \cite{Pa3} 
as function of $h$.

Dimensionless ratios   of $P(q)$ moments such as the Binder parameter and the
skewness display a non-monotonic behavior making difficult to get a
clear determination of the transition point on the AT line.  Also 
$ad~hoc$ parameters for locating replica symmetry breaking transitions, based
on the non-self-averageness of the order parameter, are considered.
The connected ones turn out to be effective for locating the transition.
 
An even better evidence for the transition comes from the divergence
of the spin glass susceptibility, though its scaling behavior is affected
by strong finite size corrections.

We also investigate the behavior of various quantities defined in term
of the probability distribution of the absolute value of the overlap.
This allows to reduce the finite size effects due to the long tail of
$P(q)$ in the negative overlap region. As a matter of fact, the
dimensionless parameters turn out to behave better in this case, the
crossing points being nearer to the correct critical value. It is
interesting to note that the usual and ``absolute'' susceptibilities
have corrections of opposite signs.

Moreover we studied the behavior of some sum rules (related to stochastic 
stability) involving overlaps between three and four replicas. We found strong
finite size corrections particularly for those sum rules that are valid only 
at non-zero magnetic field, and  it turns out to be particularly appropriate
to look at ``absolute'' quantities  in this case.
They are satisfied within our statistical accuracy for $N=1024$ in the 
glassy phase. On the other hand, they would not be good indicators for
the transition, since their behavior change very slightly when crossing
the AT line, being still nearly verified also for $h>h_{AT}$, some
trivially (all the terms become very similar) others 
non-trivially.

Finally we presented data for $P(q)$ in magnetic field, which show how
slowly the shape predicted by the RSB solution develops on a large
system.

\end{section}

\begin{section}*{Acknowledgments}
\noindent
We are particularly grateful to Giorgio Parisi for many interesting
discussions and useful suggestions. We would like to thank Tamas
Temesv\'ari for pointing to us an error in the original manuscript and for
many discussions. We also acknowledge discussions with
Andrea Crisanti, Cirano De Dominicis, Enzo Marinari, Felix Ritort,
Tommaso Rizzo, and Peter Young. B. C. is supported by a Marie Curie
(EC) fellowship (contract HPMF-CT-2001-01504). We thank Andrea
Crisanti and Tommaso Rizzo for providing us with their unpublished results on the
$h\neq 0$ SK model.
\end{section}

\end{document}